\documentclass[fleqn,usenatbib]{mnras}

\usepackage[usenames,dvipsnames]{xcolor}
\usepackage{amsmath}
\usepackage{amssymb}
\usepackage{array}
\usepackage{xfrac}
\usepackage{multirow}

\newcommand{\rev}[1]{{\textcolor{black}{#1}}}
\usepackage[T1]{fontenc}
\usepackage{ae,aecompl}
\hypersetup{draft}

\title[SFHs and abundances]{The competing effects of recent and long-term star formation histories on oxygen, nitrogen, and stellar metallicities} 

\author[Boardman et al.]{
N.~F.~Boardman$^{1}$\thanks{E-mail: nfb@st-andrews.ac.uk},
V.~Wild$^{1}$, N.~Vale Asari$^{2}$, F.~D’Eugenio$^{3,4}$\\
$^{1}$School of Physics and Astronomy, University of St Andrews, North Haugh, St Andrews KY16 9SS, UK\\
$^{2}$Departamento de F\'{\i}sica--CFM, Universidade Federal de Santa Catarina, C.P.\ 5064, 88035-972, Florian\'opolis, SC, Brazil \\
$^{3}$Kavli Institute for Cosmology, University of Cambridge, Madingley Road, Cambridge, CB3 0HA, United Kingdom \\
$^{4}$Cavendish Laboratory - Astrophysics Group, University of Cambridge, 19 JJ Thomson Avenue, Cambridge, CB3 0HE, United Kingdom
}

\date{Accepted 2025 May 13. Received 2025 May 8; in original form 2025 March 21}
\pubyear{2025}
\begin{document} 
\label{firstpage}
\pagerange{\pageref{firstpage}--\pageref{lastpage}}
\maketitle

\begin{abstract}

The fundamental metallicity relation (FMR) --- the three-way trend between galaxy stellar masses, star-formation rates (SFRs) and gaseous metallicities --- remains amongst the most studied extragalactic relations. Furthermore, metallicity correlates particularly tightly with gravitational potential. Simulations support a shared origin for these relations relating to long-term gas inflow history variations; however, differences between simulated and observed galaxy samples make it unclear whether this holds for real galaxies. We use MaNGA integral field observations to probe these relations in star-forming galaxies at one effective radius. We confirm the FMR and equivalent relations for stellar metallicity (FMR$^*$) and gaseous N/O (fundamental nitrogen relation, FNR). We find that all relations persist when considering gravitational potential in place of stellar mass and/or considering stellar ages in place of SFR, with the gaseous relations strengthened significantly by considering potential. The gaseous FMR disappears at high masses/potentials, while the FNR persists and the FMR$^*$ strengthens. Our results suggest a unified interpretation of galaxies’ gaseous and stellar metallicities and their N/O abundances in terms of their formation histories. Deeper gravitational potentials correspond to earlier star-formation histories (SFHs) and faster gas consumption, producing tight potential-abundance relations for stars and gas. In weak potentials, galaxy SFR variations primarily result from recent gas inflows, mostly affecting gas abundances. In deeper potentials, SFR variations instead correspond to broad differences in SFH shapes resulting from differences in long-term gas consumption histories, which is most visible in stellar abundances. This unified interpretation could be confirmed with upcoming higher redshift spectroscopic surveys.

\end{abstract}


\section{Introduction}\label{intro}

Stellar and gaseous chemical abundances are key observables for studying galaxy evolution. Galaxy gas abundances represent the end products of successive stellar generations, shaped also by  competing physical processes of inflow, outflow, and nucleosynthesis in stars \citep[e.g.][]{schmidt1963,lilly2013,bb2018, yang2024}. Stellar abundances meanwhile represent the history of a galaxy's gas abundances, with stars having been formed from gas at earlier times. We therefore stand to gain much insight into galaxy evolution by studying both stellar abundances and gaseous abundances \textit{together} in galaxies, an approach that has become increasingly common in recent years \citep[e.g.][]{lian2018a,frasermckelvie2022,sm2024a,stanton2024,zhuang2024,zinchenko2024}.

The gaseous metallicity of galaxies is commonly parametrised via their oxygen abundances, $\mathrm{12 + \log(O/H)}$, which can be determined from emission lines in star-forming galaxies' optical spectra. Oxygen is produced predominantly in massive ($\mathrm{>8\mathrm{M_\odot}}$) stars and then released via core-collapse supernovae (SNe) on $\sim$10 Myr timescales \citep[e.g.][]{timmes1995}, making oxygen abundances sensitive to star-formation over similarly short timescales. The gaseous metallicity of star-forming galaxies is tightly related to galaxies' stellar mass ($M_*$), in what is commonly referred to as the mass-metallicity relation (MZR). This relation has been observed both in the local Universe \citep[e.g.][]{lequeux1979,tremonti2004,andrews2013,scholte2024} and at higher redshifts \citep[e.g.][]{maiolino2008,zahid2014,jones2020}, and it has been reproduced by various simulations and models \citep[e.g.][]{derossi2017,marszewski2024}. The relation has been shown to evolve with redshift, with higher-redshift galaxies possessing lower metallicities at a given mass \citep[e.g.][]{maiolino2008}. 

Various interpretations have been suggested for the MZR. It has been suggested to be a result of increased metal-loss via outflows at low galaxy masses \cite[e.g.][]{tremonti2004,tortora2022} or higher star-formation efficiencies in higher-mass galaxies \citep{calura2009}. The MZR could also be explained as a consequence of the `downsizing' trend in which more massive galaxies form faster and earlier; in this case, the MZR would be due to more massive galaxies having converted a greater proportion of their gas to metals by the present day \citep[e.g.][]{spitoni2020}. These explanations need not be mutually exclusive: \citet{lilly2013} for instance find their model to best reproduce the observed MZR when increasing stellar masses are associated with increasing star-formation efficiencies and decreasing outflow efficiencies. 

A further metallicity trend which has captured much attention in the literature is the Fundamental Metallicity Relation (FMR): the three-way trend between star-forming galaxies' stellar masses, gas metallicities, and star-formation rates (SFRs), in which higher SFRs are associated with lower metallicities at a given $M_*$ \citep[e.g.][]{ellison2008,ll2010,mannucci2010}. The FMR has been found to be constant out to at least $z \simeq 2.5$ \citep[e.g.][]{cresci2012,cresci2019} but does appear to evolve at $z > 2.5$, with metallicities lower than what the local FMR would predict \citep{curti2023,nakajima2023,heintz2023,curti2024}. The FMR is also considerably stronger at low masses, with the relation reported to disappear or else invert amongst the higher-$M_*$ galaxies \citep[e.g.][]{mannucci2010,curti2020,lewis2024}. The amount of attention received by the FMR cannot be overstated: it has been studied at length over a wide range of observational datasets \rev{\citep[e.g.][]{sanchez2013,kashino2016,telford2016, bb2017,sanchez2017a,curti2020,dp2022,ma2024a,pistis2024}} and with a range of models and simulations \citep[e.g.][]{lilly2013,forbes2014,dave2017,derossi2017,torrey2019}, and it has been used to estimate galaxies' gas metallicity evolution from their \rev{star-formation histories \citep[SFHs;][]{peeples2013,munoz2015,greener2022}}.

The FMR is most commonly understood in terms of galaxies' gas reservoirs \citep[e.g.][]{dave2012,lilly2013,forbes2014}: variations in gas inflow rates are expected to lead to variations in SFRs due to changes in the gas supply, while also leading to metallicity variations due to metal-poor infalls diluting gas that is already present. While the FMR is not a strong trend \citep[e.g.][]{salim2014}, secondary metallicity relations with atomic and molecular gas mass are considerably stronger \citep[e.g.][]{bothwell2013,bothwell2016,brown2018}, which supports the view that metal-poor inflow variations are responsible for the FMR. Nonetheless, the FMR is far easier to observe than relations that directly involve gas masses, and so it remains a key benchmark for chemical evolution models and simulations.

Separately from the FMR, more compact galaxies have been repeatedly noted to possess metal-richer gas at a given $M_*$ in observations \citep[][]{ellison2008,deugenio2018,boardman2024a,ma2024a,sm2024} and in cosmological hydrodynamical simulations \citep[e.g.][]{sm2018a,ma2024}, with the observed trend being significantly stronger than the FMR \citep{ma2024a}. This three way trend is such that the parameter $M_*/R_e$ (hereafter $\mathit{\Phi_e}$), in which $R_e$ is a galaxy's half-light radius correlates particularly tightly with gas metallicity \citep{deugenio2018,sm2024,boardman2024a}. This relation, which we refer to hereafter as the $\mathrm{\Phi}$ZR, is tighter than both the MZR \citep[e.g.][]{deugenio2018} and FMR \citep{ma2024a} but can be expected to vary significantly with redshift; this expectation is reached by considering that star-forming galaxies are more compact at a given mass at earlier epochs \citep[e.g.][]{vdw2014} while also possessing lower gas metallicities for their mass.

The $\Phi$ZR is commonly interpreted as reflecting the importance of gravitational potential \citep[e.g.][]{deugenio2018,boardman2024a,sm2024}, with $\mathit{\Phi_e}$ treated as a proxy for a galaxy's potential depth. Such a link could conceivably arise from the connection between potential and escape velocity $V_{esc}$ \citep{deugenio2018}, with $V_{esc}$--metalliicty indeed observed in nearby galaxies on local \citep{scott2009,bb2018} and global \citep{scott2009} scales. The idea of a direct potential--metallicity connection has however been challenged on both observational and simulation-based grounds. The EAGLE simulations produce almost no relation between metallicity and escape velocity at a given stellar mass \citep{sm2018a}, suggesting that gravitational potential in itself is not what drives drives the $\Phi$ZR. \citet{baker2023a} have likewise argued that gravitational potential does not drive metallicity, based on the comparatively weak trends observed between gas metallicity and total dynamical mass in observed galaxies.

\subsection{The view from simulations}

The gaseous FMR is recovered in many large-scale cosmological hydrodynamical simulations. These include EAGLE \citep{schaye2015}, IllustrisTNG \citep{tng1,tng2,tng3,tng4,tng5,tng6,tng7} and MUFASA \citep{dave2016}. For all three of these simulation suites, the FMR appears to be shaped primarily by galaxies' gas reservoirs. Metallicity in EAGLE correlates most closely with the gas-to-stellar mass fraction \citep{derossi2017} which in turn is tightly linked to current SFR \citep{lagos2016}, with residual trends between metallicity and gas mass reported in all three mentioned simulation suites \citep{dave2017,derossi2017,torrey2019}. \citet{torrey2018} show IllustrisTNG galaxies to oscillate around equilibrium values of metallicity and SFR, with \citet{torrey2019} arguing this to be driven by periods of enrichment competing with periods of significant metal-poor inflow, producing an FMR when viewing galaxies as a population.

The FMR and $\Phi$ZR seem to possess a shared origin in simulations: more compact star-forming galaxies form more of their stars at early times, resulting in reduced current gas reservoirs, lower SFRs and higher metallicities at the time of observation. In particular, EAGLE reproduces the observation that more compact galaxies are metal-richer at a given $M_*$ while simultaneously reproducing the FMR \citep{sm2018a}. \citet{sm2018a} ascribe this to galaxies growing their size over time: more recent gas inflow events result in higher observed SFRs and lower observed metallicities, with the resulting galaxy also being larger than galaxies for which the last inflow event occurred longer ago. \citet{zenocratti2022} additionally point out that more spheroidal (and hence more compact) galaxies in EAGLE form stars more from consumption of their gas reservoirs than from inflowing gas, with disk-dominated galaxies possessing much higher metal-poor gas accretion rates below redshifts of 1. A similar picture to EAGLE emerges from IllustrisTNG: more compact disk galaxies form earlier in the TNG50 simulation, with such galaxies possessing heightened SFRs at redshifts $z > 2$ along with higher metallicities by the present day \citep{ma2024}. 

A key feature of both EAGLE and IllustrisNTG --- neccessary for the above picture to hold --- is that SFR positively correlates with half-mass radii over a wide range of stellar masses. This can for instance be seen in figure 1 of \citet[][]{sm2018a} for EAGLE and in figure 10 of \citet[][]{ma2024} for IllustrisTNG, and \citet{jf2024} similarly report that EAGLE galaxies possess lower sSFRs at higher concentrations. Such behaviour is however \textit{not} supported by observations of the nearby Universe, in which high SFRs are associated with smaller sizes over a significant stellar mass range \rev{\citep{wuyts2011}}. The true \rev{origins} of the $\Phi$ZR and FMR are therefore unclear, and further insight can be gained by considering additional abundance measures.

\subsection{Alternative abundance tracers: stellar metallicity}

As with the gaseous MZR, a positive relation is found between galaxy stellar masses and galaxy stellar metallicities \citep[e.g.][]{gallazzi2005}. Stellar metallicities are often parameterised in terms of the iron abundance relative to solar ($\mathrm{[Fe/H]}$); this is due to the relative ease of determining iron abundances from optical absorption-line spectroscopy \citep[e.g.][]{conroy2013}, compared to the much harder-to-measure oxygen abundance \citep[e.g.][]{asplund2004,asplund2009,ting2017,ting2018}. However, the timescales of iron and oxygen enrichment are also different, with oxygen produced \rev{almost entirely} by type-II SNe (thus during star-formation episodes) while iron is produced and released by both type-II and type-Ia supernovae, the latter of which explode over both short and long (~Gyr) timescales \citep{maoz2010}. Thus, iron enrichment is expected to continue for over a Gyr after any given star-formation episode \citep[e.g.][and references therein]{maiolino2019,johnson2021}.

Amongst star-forming galaxies, gas metallicities are typically higher than stellar metallicities \citep[][]{gallazzi2005,lian2018a} when normalised relative to solar abundances, with the size of gas--stellar metallicity offsets appearing to relate to galaxies' \rev{SFHs \citep[][]{frasermckelvie2022,zhuang2024}}. As with gas metallicities, compact galaxies possess higher stellar metallicities at a given $M_*$ \citep[e.g.][]{cappellari2013a,scott2017,li2018}. Stellar metallicities have been reported to correlate more tightly with $\mathit{\Phi_e}$ than with $M_*$, for both quiescent \citep{barone2018} and star-forming \citep{barone2020} galaxies; we refer to the stellar $\mathit{\Phi_e}$-metallicity relation as the $\Phi$ZR$^*$ for the rest of this article. This relation is equivalent to the relation between stellar metallicity and velocity dispersion, which has been repeatedly found for nearby quiescent galaxies \citep[e.g.][]{gallazzi2006,mcdermid2015}.
 
A stellar equivalent to the FMR --- between $M_*$, SFR and stellar metallicity --- has recently been reported in nearby ($z \lesssim 0.15$) massive galaxies \citep{looser2024}, building upon past reports of a stellar metallicity offset between star-forming and quiescent galaxies \citep[e.g.][]{peng2015,trussler2020}. We refer to this relation as the FMR$^*$ for the remainder of the article. A FMR$^*$ also seems to exist within low-mass star-forming galaxies at low redshift \citep{zhuang2024}, with a FMR$^*$ also reported in high-redshift ($z \sim 5$) galaxies from UV-based metallicities \citep{faisst2016}, though contrasting behaviour is detected among $0.6<z<1$ star-forming galaxies in the  LEGA-C dataset \citep{nersesian2025}. \citet{looser2024} further report that the FMR$^*$ persists when $M_*$ is replaced with $\mathit{\Phi_e}$. \citet{looser2024} argue the existence of the FMR$^*$ to imply that both FMRs are driven by long-lasting metal-poor inflows, as opposed to the FMR being driven by short-lived gas accretion events alone. A FMR$^*$ is detected within various cosmological hydrodynamical simulations \citep[Illustris, IllustrisTNG and EAGLE;][]{derossi2018,garcia2024a,looser2024} which likewise suggest it to be driven by long-term gas inflow variations.

Notably, \citet{looser2024} performed their study using Mapping Nearby Galaxies at Apache Point Observatory \citep[MaNGA;][]{bundy2015}, an integral-field unit (IFU) spectroscopic survey. The FMR has been repeatedly noted to be weak in IFU surveys, to the point of some studies questioning its existence \citep{sanchez2013, sanchez2017a, bb2017}, though see \citet{cresci2019} for a different view. The FMR has also been found repeatedly to disappear at the highest stellar masses, which is not the case for the FMR$^*$ \citep{looser2024}. This suggests that the FMR$^*$ might be \textit{tighter} than the FMR, which would in turn suggest the FMR$^*$ to be more informative on galaxy evolution.

\subsection{Alternative abundance tracers: gaseous nitrogen}

An additional relevant property is the gaseous N/O abundance ratio. Nitrogen enrichment takes place over longer timescales than oxygen enrichment \citep[e.g.][]{edmunds1978}; the enrichment appears to proceed via multiple channels, with both a metallicity-dependent and metallicity-independent component \citep[e.g.][]{edmunds1978,matteucci1986}. It is generally assumed \citep[e.g.][]{chiappini2005, vincenzo2016, johnson2023} that primary nitrogen is produced by massive ($\mathrm{>8 \mathrm{M_\odot}}$) stars on $\sim$10 Myr timescales along with oxygen, with the elements then ejected into the interstellar medium (ISM) via core-collapse supernovae at a roughly fixed fraction with oxygen. Secondary nitrogen is then produced later in this scenario in asymptotic giant branch (AGB) stars via the CNO cycle (at a metallicity-dependent rate), and this comes to dominate nitrogen production at higher metallicities. These assumptions are far from universal, however; \citet{molla2006} for instance propose a model in which massive stars instead produce secondary nitrogen, with primary nitrogen contributed by low-to-intermediate mass stars.  

Observationally in the local Universe, N/O is tightly correlated with O/H at higher metallicities in the local Universe while instead presenting a roughly fixed plateau at lower metallicities \citep[e.g.][]{chiappini2005,andrews2013,nicholls2017}. N/O correlates significantly with $M_*$ \citep[e.g.][]{pm2013,andrews2013}, with evidence of substantial redshift-evolution as with the MZR \citep{strom2017,hp2022}. N/O correlates with $\mathit{\Phi_e}$ even more tightly than O/H does \citep{boardman2024a}, and we refer to the $\mathit{\Phi_e}$--N/O relation as the $\Phi$NR hereafter. 

Gaseous N/O is expected to be much less sensitive to recent pristine gas inflows, due to it being insensitive to hydrogen content; this makes N/O a strong indicator of a galaxy’s
overall evolutionary state. It is therefore especially interesting that a `fundamental nitrogen relation' (FNR) appears to exist in nearby galaxies \citep{hp2022}, in which N/O negatively correlates with SFR at a given $M_*$; such a result can also be inferred from \citet{telford2016} and \citet{sanders2018}, who both investigate the FMR using metallicity indicators sensitive primarily to N/O. 

An FNR is also found within EAGLE \citep{matthee2018} and is ascribed to broad variations in galaxies' star-formation histories (SFHs) which, in turn, correspond to variations in galaxies' present day SFRs at fixed $M_*$ \citep{matthee2019}. Given the connection between SFHs and gas inflow histories, this suggests the FNR to have similar origins to the FMR$^*$: the FNR is driven primarily by long-lasting metal-poor inflows which fuel episodes of enhanced star-formation, as opposed to the short-term accretion events argued for the FMR.

\subsection{Towards a unified perspective of the FMR, FNR and FMR$^*$}

So far, the three mass-SFR-abundance relations -- the FMR, FMR$^*$ and FNR -- have mostly been considered separately in the literature. Thus, a detailed census of all three relations \textit{together} is clearly called for. For instance, how strong are the FMR$^*$ and FNR at different masses, and how do these compare with the traditional FMR? How do the strengths of the three relations compare when $M_*$ is replaced with $\mathit{\Phi_e}$, and what can this tell us about the physics behind chemical evolution? How does the stellar--gas metallicity offset vary as a function of $M_*$ and SFR, or of $\mathit{\Phi_e}$ and SFR, given the apparent connection between stellar-gas metallicity offset and SFH? The answers to these questions will provide important benchmarks for chemical evolution models, which will then inform the processes responsible for chemical enrichment in galaxies. 

With the above questions in mind, we consider the FMR, FNR and FMR$^*$ together for the first time, using spectroscopic data from the MaNGA survey. We compare the strength of these three relations, considering both $\mathit{\Phi_e}$ and $M_*$. We also consider the importance of past SFH more directly by using stellar population ages as an alternative to SFR, \rev{with younger stellar populations having previously been reported to correspond to lower metallicities at fixed $M_*$ \citep{lian2015,sm2020}}. We describe our sample and data in \autoref{sampledata}, and we present our results in \autoref{results}. We discuss our findings and then conclude in \autoref{disc}. We assume a \citet{chabrier2003} initial mass function and adopt the following $\Lambda$ Cold Dark Matter cosmology: $\mathrm{H_0} = 71$ km/s/Mpc, $\mathrm{\Omega_M} = 0.27$, $\mathrm{\Omega_\Lambda} = 0.73$.

\section{Sample \& data}\label{sampledata}

We obtain our galaxy sample and much of our data from the SDSS-IV \citep{blanton2017} MaNGA survey \citep{bundy2015}, for which all data and analysis products are publically available as of SDSS-IV Data Release 17 \citep[DR17;][]{sdssdr17}. MaNGA performed integral field unit (IFU) spectroscopy on just over 10000 nearby ($z \lesssim 0.15$) galaxies selected for a wide range of morphologies and a roughly flat distribution in $\log(M_*)$\citep{yan2016b,wake2017}. The MaNGA instrument employs a series of hexagonal fibre bundle IFUs with fibre configurations ranging from 19 to 127 fibres \citep{drory2015}, with observations taken using the BOSS spectrographs on the 2.5~m Sloan telescope at Apache Point Observatory \citep{gunn2006,smee2013}. A three-point dithering pattern was employed in the observations, to fully sample the field of view \citep{law2015}. The observations were reduced through the MaNGA data reduction pipeline \citep[DRP][]{yan2016a, law2016} and then analysed via the MaNGA data analysis pipeline \citep[DAP;][]{belfiore2019,westfall2019, law2021}.

Our data collection procedures mostly follow \citet{boardman2024a}.  We obtained elliptical Petrosian axis ratio $b/a$, $M_*$, and $r$-band elliptical Petrosian $R_e$ values from the NASA-Sloan-Atlas (NSA) catalog \citep{blanton2011}\footnote{\url{https://data.sdss.org/sas/dr17/sdss/atlas/v1/nsa_v1_0_1.fits}}.  We obtained a parent sample by selecting MaNGA galaxies in the Primary+ and Secondary samples, which respectively were observed out to $\sim$ 1.5\,$R_e$ and $\sim$ 2.5\,$R_e$, and by then restricting to galaxies with $b/a > 0.6$ to avoid edge-on systems. 

We assumed 0.1 dex uncertainties on stellar masses, representative of stellar mass uncertainties in the MPA-JHU catalog\footnote{\url{https://www.sdss4.org/dr17/spectro/galaxy_mpajhu/.} \citep[see also][]{frasermckelvie2019b}}, and 0.05 dex uncertainties on $R_e$ \citep{deugenio2018}. Our adopted stellar mass errors are broadly representative but should be viewed as conservative. If we select all MPA-JHU galaxies with $\log(M_*/\mathrm{M_\odot}) > 8 $ and $sSFR > -11)$, we find a median mass error of 0.095 dex. A mild trend with mass is evident in the errors: if we restrict these star-forming MPA-JHU galaxies to a mass range $ 8.9 < \log(M_*/\mathrm{M_\odot}) < 9.1 $, we obtain a median mass error of 0.082 dex, whereas for a mass range $10.9 < \log(M_*/\mathrm{M_\odot}) < 11.1$ we obtain a median error of 0.10 dex. Thus, an adopted error of 0.1 dex is appropriate for our purposes. 

We obtained emission line flux maps and associated errors from the DAP for $\mathrm{H~\alpha}$, $\mathrm{H~\beta}$, [O~\textsc{iii}]$_{5008}$, [N~\textsc{ii}]$_{6585}$, [S~\textsc{ii}]$_{6718}$, [S~\textsc{ii}]$_{6733}$ and [O~\textsc{ii}]$_{3737, 3729}$. We also obtained $\mathrm{H~\alpha}$ equivalent widths $EW_{H\alpha}$ \rev{from the DAP}. We used non-parametric summed emission fluxes in all cases, following subtraction of the stellar continuum. We restricted to spaxels with signal-to-noise ratios $S/N > 3$ for all emission line fluxes. The DAP emission values are corrected for Milky Way foreground extinction using the \citet{schlegel1998} maps with a \citet{odonnell1994} dust law, though they are not corrected for galactic dust. We corrected emission lines for dust using a \citet{fitzpatrick2019} correction curve, assuming an intrinsic Balmer decrement $\mathrm{H~\alpha/H~\beta} = 2.86$.

As described in \citet{boardman2024a}, we derived maps of star-forming spaxels' gaseous metallicity and N/O using the \citet{curti2020} RS32\footnote{Defined as $\mathrm{[S~\textsc{ii}]_{6718,6733} / H~\alpha + [O~\textsc{iii}]_{5008} / H~\beta}$.} calibrator and the \citet{florido2022} N2O2 calibrator. We selected star-forming spaxels using BPT diagnostics \citep{bpt,osterbrock1985,veilleux1987}, specifically BPT--NII \citep{kauffmann2003} and BPT--SII \citep{kewley2001}, and by requiring $EW_{H\alpha} > 14$\AA\ \citep{lacerda2018,valeasari2019}. We then calculated gas abundance values at 1\,$R_e$, $\mathrm{12 + \log(O/H)_e}$ and $\mathrm{\log(N/O)_e}$, along with associated errors. Our final sample consists of all galaxies for which reliable gas abundances could be obtained, which results in 2070 star-forming galaxies. Such a sample selection is necessary for studying gas abundances, though we note that stellar metallicities could also be studied in quiescent galaxies; we choose to restrict to the same 2070 galaxies when studying stellar metallicities, since we specifically wish to study the behavior of different abundance measures \textit{together} in nearby galaxies.

We obtained SFRs from the DR17 pyPipe3d analysis release \citep{sanchez2022}, which used an improved version of the Pipe3d pipeline \citep{sanchez2016,sanchez2016b,sanchez2019}. We use here the \textit{log\_SFR\_Ha} column from the pipe3d summary file, which is the integrated SFR determined by summing all detected $\mathrm{H~\alpha}$ emission within the MaNGA field of view. This SFR should be viewed as an upper limit, since it does not take different sources of ionization into account; however, we have verified that we obtain effectively identical results if we use \textit{log\_SFR\_SF} (which uses $\mathrm{H~\alpha}$ calculations for star-forming spaxels only). We also obtain similar results with \textit{log\_SFR\_ssp\_10Myr}, which provides SFR estimations over 10 Myr from full spectral fitting of the stellar continuum, though we caution that spectral fitting over optical wavelenghs alone can  produce overestimations in the recent SFR \citep{lopezfernandez2016,werle2019}.

When comparing observational results to simulations, an important caveat lies in the size measures used in different studies. Simulation-based studies typically use half-\textit{mass} radii \citep[e.g.][]{sm2018a,ma2024}, whereas observational studies --- including this work and the previously-mentioned \citet{wuyts2011} --- overwhelmingly use light-based size parameters. If compact star-forming galaxies emerge from recent starburst events \citep{wuyts2011}, then one might expect their half-mass radii and half-light radii to behave differently as a function of stellar mass \citep{gb2017}, though this would remain inconsistent with the simulation-based view of more compact galaxies having formed earlier. Thus, we obtain half-mass radii $\mathrm{R50_*}$ from the \textit{R50\_kpc\_mass} column of the MaNGA pipe3d summary catalog \citep{sanchez2022}, for which the radii were computed over the MaNGA field of view. 

\begin{figure}
\begin{center}
	\includegraphics[trim = 0.8cm 1.8cm 1cm 10.3cm,scale=1,clip]{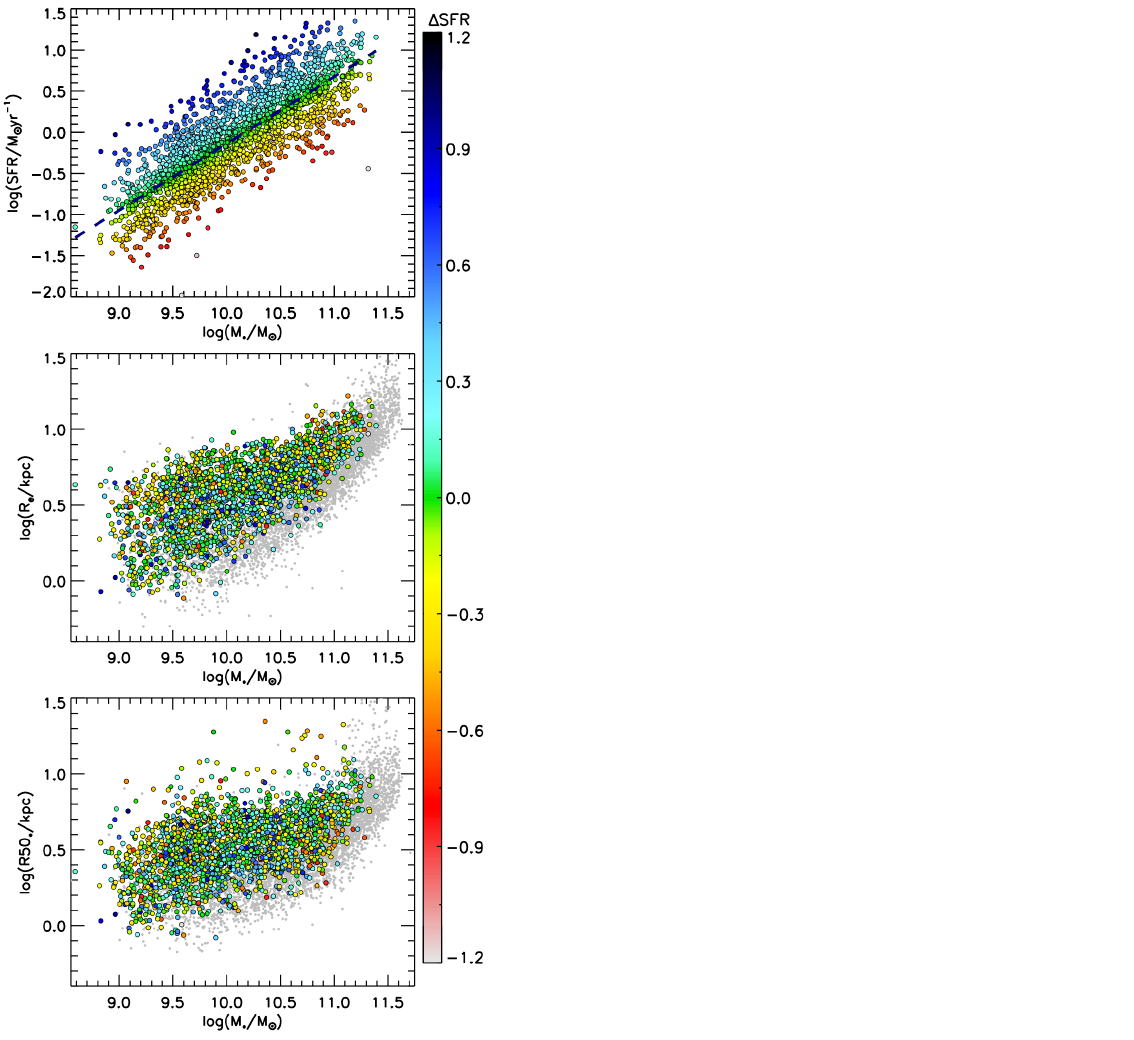} 
	\caption{Top: star-forming sequence (SFS) for our final star-forming galaxy sample. The dashed line shows the straight-line fit to the SFS as described in the text. Middle: mass-size plane for our final sample (colored points) and parent sample (grey points), using half-light radii. Bottom: mass-size plane using half-mass radii. The star-forming galaxy sample is colored by $\mathrm{\Delta SFR}$, which indicates the difference between galaxies' SFRs and the fitted SFS.}
	\label{sampleplot}
	\end{center}
\end{figure}

We present our final sample in terms of $M_*$, SFR, $R_e$ and $\mathrm{R50_*}$ in \autoref{sampleplot}, in which a tight star-forming sequence (SFS) is evident. We perform a straight line fit to the SFS using the \textsc{IDL} \rev{\textsc{LADFIT}} routine, and we compute the offset $\mathrm{\Delta SFR}$ of individual galaxies' SFRs from this fit. Positive values of $\mathrm{\Delta SFR}$ indicate galaxies with high SFRs for their mass, as shown by the color scale in \autoref{sampleplot}.

\begin{figure*}
\begin{center}
	\includegraphics[trim = 0.2cm 2.4cm 0cm 15.5cm,scale=1,clip]{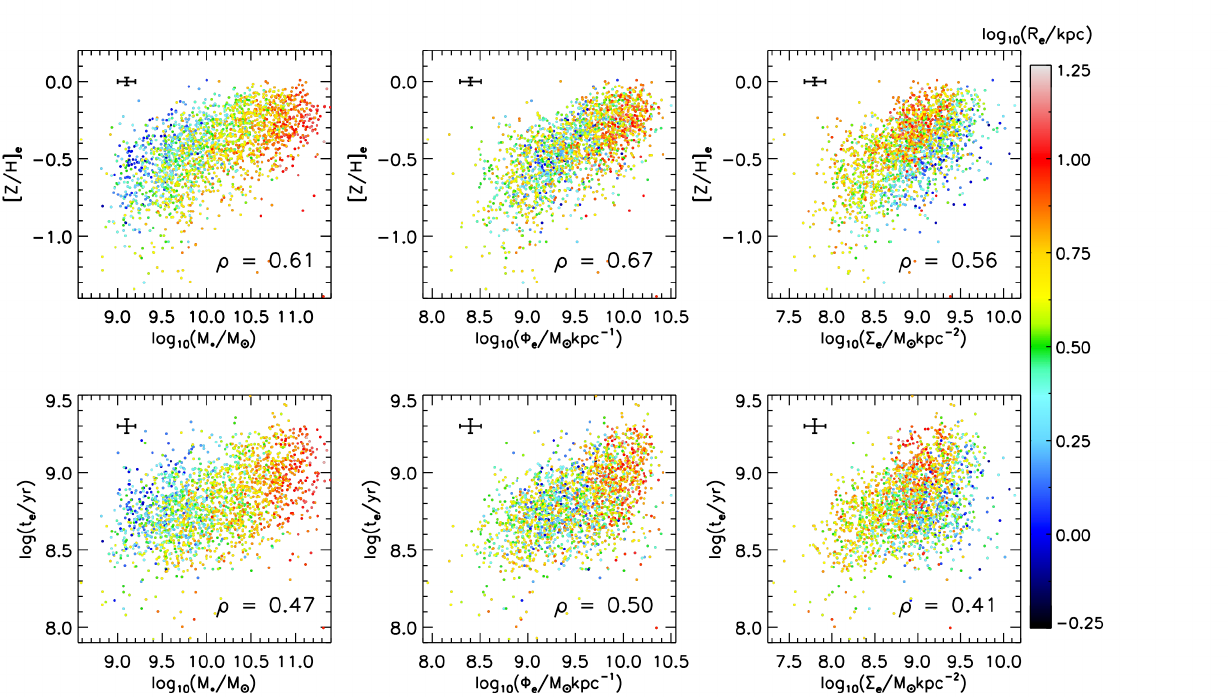} 
	\caption{Light-weighted pipe3d stellar metallicities (top) and stellar ages (bottom) at $\mathrm{1R_e}$, plotted against $M_*$ (left panels), $\mathit{\Phi_e}$ (middle panels) and $\mathit{\Sigma_e}$ (right panels). Each panel displays the corresponding Spearman correlation coefficient ($\rho$), with $\mathrm{P} << 0.01$ in all cases. We color data points by galaxies' $R_e$, with error bars showing the median uncertainties.}
	\label{age_colorre}
	\end{center}
\end{figure*}

From comparing the middle and bottom panels of \autoref{sampleplot}, we find that $R_e$ and $\mathrm{R50_*}$ both show little relationship with $\mathrm{\Delta SFR}$ over much of the sample's mass range, though we note a mild positive $\mathrm{R50_*}$--$\mathrm{\Delta SFR}$ trend at the high-mass end of our sample; we confirmed this feature by applying locally weighted regression (LOESS) smoothing \citep{cleveland1988} to the middle and bottom panels, with the smoothing not revealing any other significant differences in SFR behavior. We also find $\mathrm{R50_*}$ to be well-correlated with $R_e$ for our star-forming galaxy sample (Spearman correlation $\rho = 0.86$, with $P \ll 0.01$). We thus find half-mass and half-light radii to behave similarly in our sample, meaning that switching to half-mass radii would \textit{not} notably decrease tension with simulations concerning galaxies' size--SFR behavior.

We additionally obtained light-weighted stellar ages $t_e$ and metallicities $\mathrm{[Z/H]_e}$ calculated at $1R_e$ from the DR17 pyPipe3d release, with the ``e" subscript indicating values at $1R_e$ throughout this work. Pipe3d performs spatial binning of spaxels to achieve higher-S/N continuum spectra while preserving a given galaxy's spatial shape as much as possible. It fits the stellar component of spectra using a simple stellar population (SSP) library built from the MaStar stellar library \citep{yan2019}, with the SSP library consisting of 273 spectra sampling 39 ages (1 Myr to 13.5 Gyr) and 7 metallicities ($Z/Z_\odot = $0.006, 0.029, 0.118, 0.471, 1, 1.764, 2.353).  Pipe3d includes dust attenuation in its fits via the \citet{cardelli1989} extinction law with $R_\mathrm{V} = 3.1$. Pipe3d then determines $1R_e$ values for age and metallicity by grouping MaNGA stellar population maps into annuli and then performing linear regression straight-line fits between galactocentric radii of 0.5 and 2 $R_e$.

Pipe3d's adopted SSP library assumes solar abundances, meaning that its derived stellar metallicities ($\mathrm{[Z/H]}$) can be considered functionally equivalent to iron abundances ($\mathrm{[Fe/H]}$). The stellar metallicity is constrained mostly by iron absorption features when performing full spectral fits over optical wavelengths \citep[e.g.][]{leung2024}, further cementing its equivalence to $\mathrm{[Fe/H]}$. This means that gas and stellar metallicities are in different elemental bases, with gas metallicities traced by oxygen and stellar metallicities instead traced by iron. Given the substantially different timescales over which oxygen and iron enrichment take place, this is an important caveat when directly comparing stellar and gaseous metallicities in galaxies \citep[e.g.][]{frasermckelvie2022}.

\section{Results}\label{results}

In this section we will consider both mass-abundance relations and $\mathit{\Phi_e}$-abundance relations, with abundances measured at $1R_e$ in all cases, to study residual abundance trends with $\mathrm{\Delta SFR}$ and with light-weighted ages $t_e$. Firstly however, it is worth considering how stellar metallicities $\mathrm{[Z/H]_e}$, $\log(t_e)$  and $\mathrm{\Delta SFR}$ vary with galaxies' stellar masses and sizes. We therefore present such an analysis in \autoref{sec31}, while referring readers to \citet{boardman2024a} for a similar MaNGA analysis of gaseous O/H and N/O. We present visualisations of key abundance relationships in \autoref{sec32}, before comparing the strengths of different relationships in \autoref{sec33}. Finally, we assess offsets between gas and stellar metallicities in \autoref{sec34}.

\subsection{[Z/H], $t_e$ and SFR as functions of mass and size}\label{sec31}

\begin{figure*}
\begin{center}
	\includegraphics[trim = 2cm 1cm 1cm 9cm,scale=0.95]{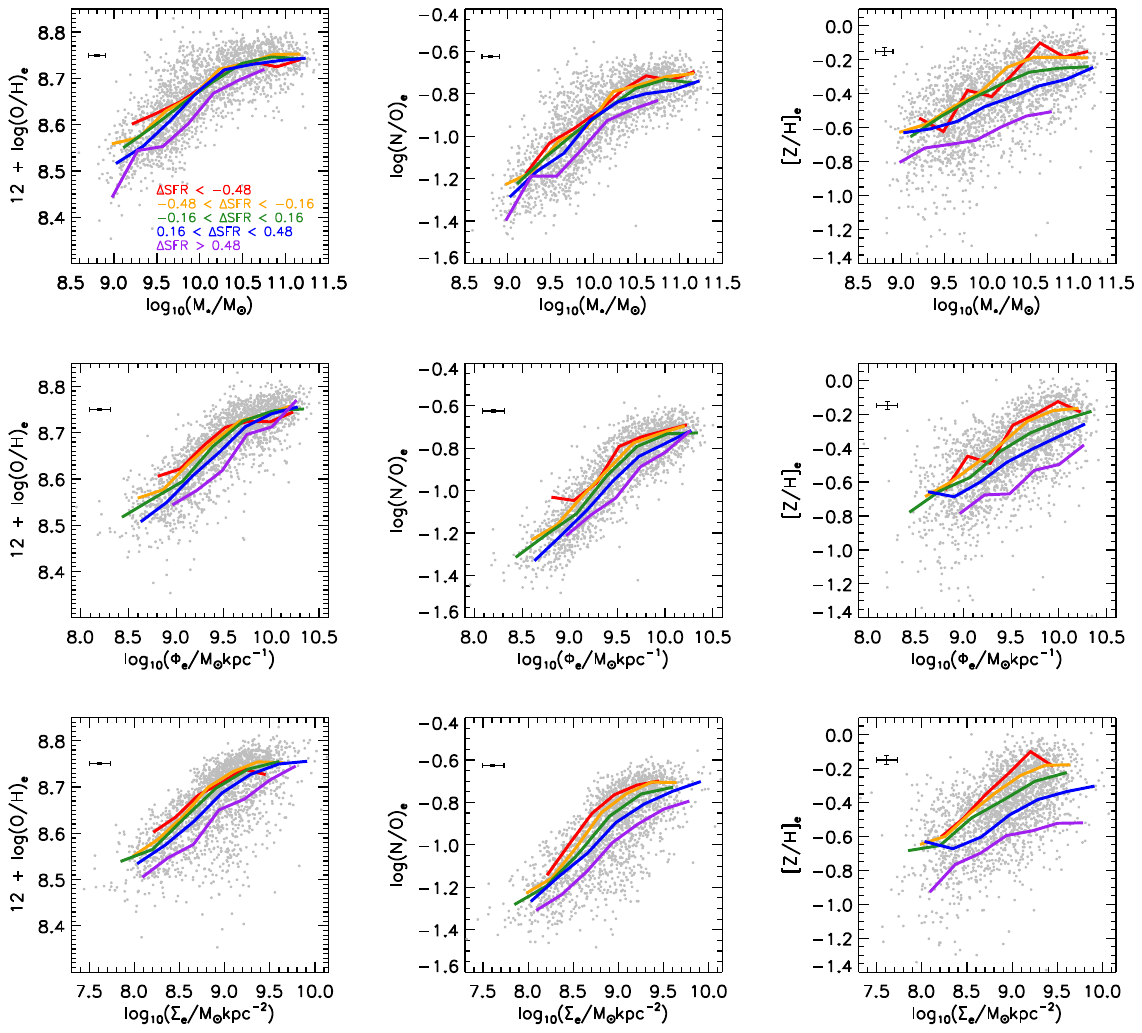} 
	\caption{Mass--SFR--abundance (top row; FMR left, FNR middle, FMR$^*$ right), potential--SFR--abundance (middle; $\Phi$-FMR left, $\Phi$-FNR middle, $\Phi$-FMR$^*$ right) and density--SFR--abundance (bottom row) relations. We show relations for gas metallicity (left column), gaseous N/O (middle column) and stellar metallicity (right column), with all abundances measured at 1 half-light radius. The colored lines show medians in bins of $M_*$ (top),$\mathit{\Phi_e}$ (middle) or $\mathit{\Sigma_e}$ (bottom), for different ranges of $\mathrm{\Delta SFR}$, while grey point represent individual galaxies. We find a residual SFR--abundance trend in all cases, with more star-forming galaxies having lower abundances. The black error bars indicate median uncertainties.}
	\label{fmr_overview}
	\end{center}
\end{figure*}

We plot in \autoref{age_colorre} galaxies' stellar metallicity and age as a function of $M_*$, $\mathit{\Phi_e}$ and stellar density $M_*/R_e^2$ (hereafter $\mathit{\Sigma_e}$) in turn; this latter parameter was reported to be the most predictive of stellar age out of the three in \citet{barone2020}. We show the corresponding Spearman rank correlation coefficient $\rho$ in each case; going forward, all reported $\rho$ values have corresponding p-values of $P < 0.01$ unless otherwise stated. We color data points by $\log(R_e/\mathrm{kpc})$ in this figure. For conciseness, we do not plot $\mathrm{\Delta SFR}$, but we describe results for this parameter below.

\textit{We find $\mathit{\Phi_e}$ to produce the strongest age correlation out of the three considered parameters}, contrary to the results of \citet{barone2020}. The cause of this difference is unclear, with differences in sample and methodology both being possible factors; in particular, we note that \citet{barone2020} used light-weighted stellar ages calculated \textit{inside} apertures of radius 1~$R_e$, while we use point estimates calculated \textit{at} 1~$R_e$. We find a clear residual $R_e$ trend when considering $\log(t_e)$ and $\mathit{\Sigma_e}$ together --- $t_e$ positively correlates with $R_e$ by eye (\autoref{age_colorre}, bottom right panel) --- whereas no such trend is apparent in the $t_e$--$\mathit{\Phi_e}$ relation (\autoref{age_colorre}, bottom middle panel); this signifies that the weaker correlation between $t_e$ and $\mathit{\Sigma_e}$ is not simply due to increased errors. $\mathrm{\Delta SFR}$, meanwhile, displays only mild correlations with $\mathit{\Phi_e}$ or $\mathit{\Sigma_e}$ ($\rho = 0.074$ and $\rho = 0.16$ respectively) while displaying no statistically significant correlation with $M_*$ by construction ($\rho = 0.007$, $P = 0.74$). Finally, and in agreement with \citet{barone2020}, we find from \autoref{age_colorre} that $\mathit{\Phi_e}$ produces a stronger correlation with $[Z/H]_e$ when compared to $M_*$ or $\mathit{\Sigma_e}$. 

\subsection{The `fundamental' relations}\label{sec32}

We plot in \autoref{fmr_overview} the FMR, FNR and FMR$^*$ for our sample. In the same plot, we also show alternative versions of these relations in which $M_*$ is replaced with $\mathit{\Phi_e}$ or with $\mathit{\Sigma_e}$. We detect three-way relations in all nine cases, showing that \textit{galactic abundances are inherently connected to SFR even after accounting for both stellar mass and size}. This is in agreement with what \citet{looser2024} report for MaNGA stellar metallicities in the case of $M_*$ and $\mathit{\Phi_e}$ but is different from \citet{vaughan2022}, who find from SAMI spectroscopy that $\mathit{\Phi_e}$ largely eliminates residual SFR trends in stellar metallicities\rev{;} we discuss this disagreement further in \autoref{disc_obs}. \rev{Our findings are also consistent with what \citet{ma2024a} report for MaNGA gas metallicities.}

\begin{figure*}
\begin{center}
	\includegraphics[trim = 2cm 1cm 1cm 9cm,scale=0.95]{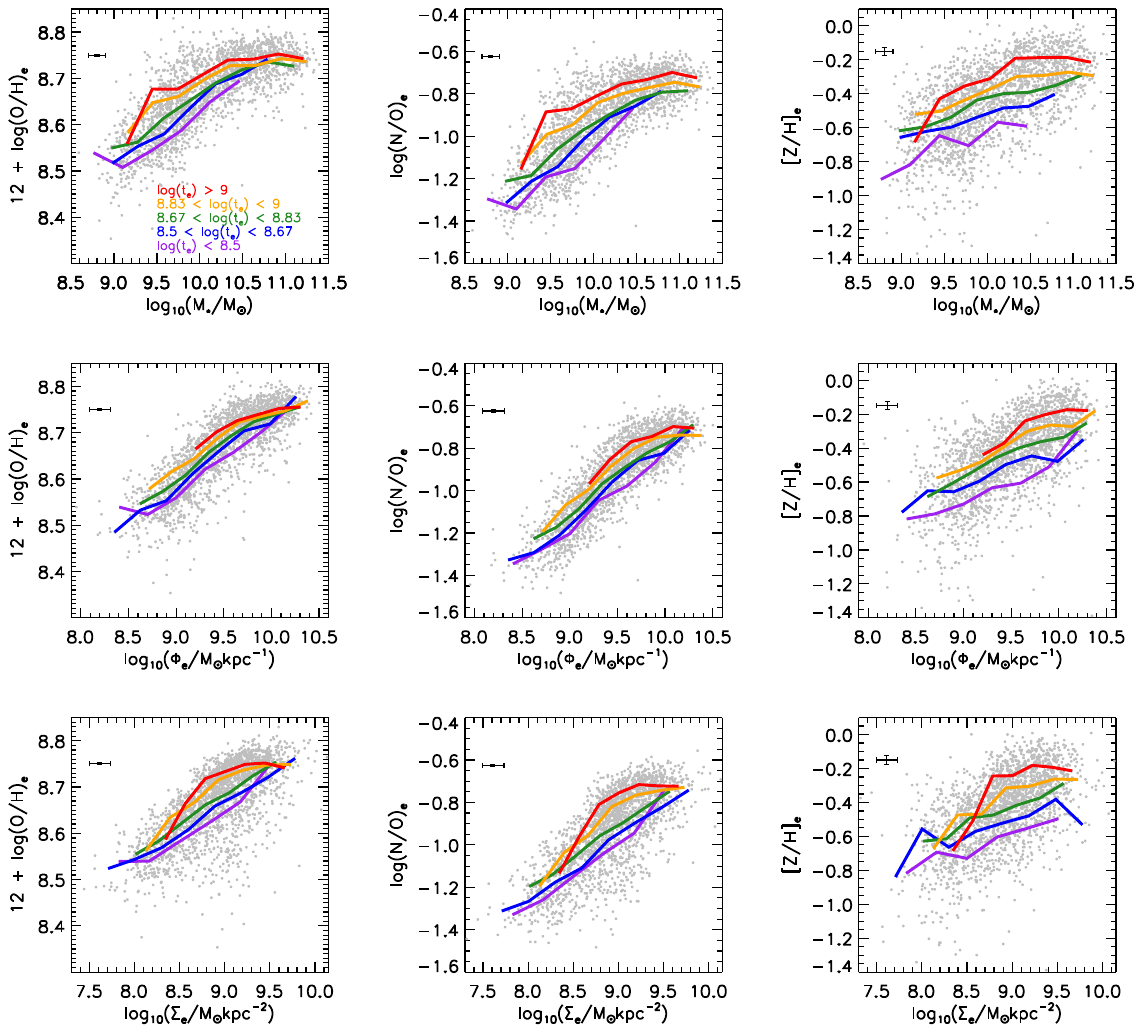} 
	\caption{Mass--age--abundance (top), potential--age--abundance (middle) and density--age--abundance (bottom) relations for gas metallicity (left), gaseous N/O (middle) and stellar metallicity (right), with all abundances measured at 1 half-light radius. The format is as in \autoref{fmr_overview}. We find a residual age--abundance trend in all cases, with younger galaxies found to prefer lower abundance values.}
	\label{fmr_overview_te}
	\end{center}
\end{figure*}

In \autoref{fmr_overview_te} we present relations in which $\mathrm{\Delta SFR}$ has been replaced with $\log(t_e)$, using the same format as for \autoref{fmr_overview}. We detect a three-way trend in all nine cases as before, \rev{with older ages associated with higher metallicities as expected; a mass-age-O/H relation, we note, has been demonstrated before with MaNGA pipe3d results by \citet{sm2020}}. By visually comparing the top left and top middle panels of the two figures, we see that $t_e$ yields rather more pronounced gas abundance trends than does $\mathrm{\Delta SFR}$ when $M_*$ specifically is considered; in other panels the residual trends appear comparable with what $\mathrm{\Delta SFR}$ yields.

For the remaining subsections, we specifically consider residual trends involving mass-abundance and $\mathit{\Phi_e}$-abundance relations. We consider $\mathit{\Sigma_e}$-abundance relations to be less informative, due to the weaker correlations seen in \autoref{age_colorre} and in \citet{boardman2024a} for gaseous N/O$\mathrm{_e}$ and O/H$\mathrm{_e}$. For the remainder of this article, we refer to the $\mathit{\Phi_e}$--SFR--abundance relations as the $\Phi$-FMR, $\Phi$-FNR and $\Phi$-FMR$^*$, analogous to the three mass-based `fundamental' relations.

\subsection{How informative are different `fundamental' relations?}\label{sec33}

\begin{figure*}
\begin{center}
	\includegraphics[trim = 0.4cm 0.3cm 1cm 2.2cm,scale=0.8]{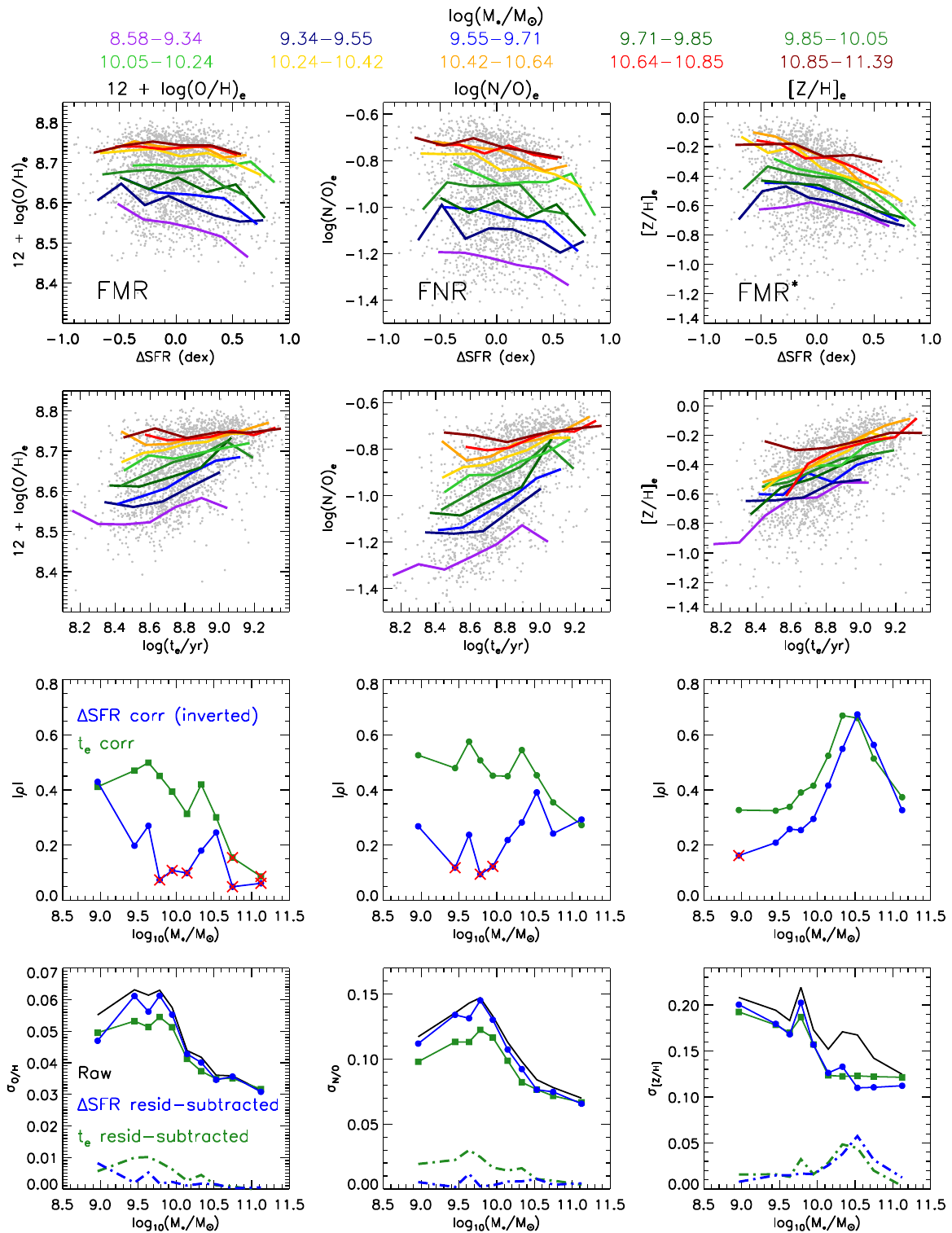} 
	\caption{\textit{Top}: the FMR, FNR and FMR$^*$. The colored lines show median abundances as a function of $\mathrm{\Delta SFR}$, calculated in ten bins of stellar mass each comprising $\sim$10 \%\ of sample galaxies. We show results for the FMR (left), FNR (middle) FMR$^*$ (right) in turn. \textit{Top middle}: equivalent relations involving stellar age, in the same format as the top panels. \textit{Bottom middle}: Spearman rank correlation coefficients $\rho$ between $\mathrm{\Delta SFR}$ (or $t_{e}$) and abundance measures, in bins of stellar mass. Red Xs indicating statistically insignificant correlations ($P > 0.01$). We show the absolute value of the coefficients for easier comparisons. \textit{Bottom}: dispersion in abundance measures before (black lines) and after subtracting out fitted $\mathrm{\Delta SFR}$--abundance relations (blue circles) or fitted age--abundance relations (green squares); the colored \rev{dot-dashed}  lines at the bottom indicate the differences in scatter before/after subtracting the fitted relations.} 
	\label{fmr_assess}
	\end{center}
\end{figure*}

We now characterize the strength of the various `fundamental' relations involving $M_*$ or $\mathit{\Phi_e}$. We proceed by splitting our sample in ten $M_*$ (or $\mathit{\Phi_e}$) bins which each comprise $\sim$10\%\ of galaxies. In each bin, we calculate the Spearman correlation coefficient $\rho$ between a given abundance measure and $\mathrm{\Delta SFR}$ (or $t_e$), to track how the strengths of residual trends vary as functions of $M_*$ or of $\mathit{\Phi_e}$. This approach does not consider residual abundance correlations with $M_*$ ($\mathit{\Phi_e}$) within a given $M_*$ ($\mathit{\Phi_e}$) bin. However, we have verified that partial correlation coefficients controlling for $M_*$ ($\mathit{\Phi_e}$) within a given $M_*$ ($\mathit{\Phi_e}$) bin yield entirely equivalent results, so we show Spearman correlation coefficients for simplicity. 

We also calculate the dispersion ($\sigma$) of abundance measures in each bin, using the \textsc{IDL} routine \textsc{ROBUST\_SIGMA}. We perform these calculations before and after accounting for residual trends in $\mathrm{\Delta SFR}$ or $t_e$, to assess how these parameters contribute to the scatter. For a given abundance measure in a given mass or potential bin, we perform these calculations in the following manner:

\begin{enumerate}
    \item Calculate an initial $\sigma$ for data points within the bin.
    \item Fit a third-order polynomial to abundances within the bin, as a function of a third parameter of interest ($\mathrm{\Delta SFR}$ or $t_e$).
    \item Subtract the polynomial out from the abundance measures.
    \item Calculate a new $\sigma$ from the polynomial-subtracted abundances. This captures all sources of dispersion \textit{minus that associated with the adopted third parameter.}
    
\end{enumerate}

In the top panels of \autoref{fmr_assess}, we plot the three considered abundance measures as a function of $\mathrm{\Delta SFR}$ in bins of stellar mass; this provides alternative visualisations of the FMR, FNR and FMR$^*$. Below these panels, we plot abundances as functions of $\log(t_e)$  for the same stellar mass bins. We also show in \autoref{fmr_assess} the Spearman correlation coefficients and dispersions for each abundance measure in the same mass bins, before and after accounting for $\mathrm{\Delta SFR}$ or $\log(t_e)$  as a third parameter. We show $\rho$ values in the third row of panels from top, with dispersions $\sigma$ shown in the bottom row. 

In \autoref{fmr_assess}, higher correlation coefficients indicate that $\mathrm{\Delta SFR}$ or $t_e$ are more informative of an abundance measure within a given mass bin. Similarly, larger $\sigma$ reductions between the initial $\sigma$ (black lines) and the $\sigma$ corrected for $\mathrm{\Delta SFR}$ (blue lines) or $\log(t_e)$ (green lines) indicate a greater reduction in scatter, which likewise indicate $\mathrm{\Delta SFR}$ or $\log(t_e)$ to be more informative on an abundance measure. From visual inspection, higher $\rho$ values are typically associated with greater $\sigma$ reductions as expected. The absolute values of $\sigma$ will depend on the range of the abundance measure being plotted, so $\sigma$ values should not be compared across different abundance measures in \autoref{fmr_assess}; instead, one should concentrate on the relative $\sigma$ differences within a given panel, which are shown at the bottom of each panel as coloured lines.  

\begin{figure*}
\begin{center}
	\includegraphics[trim = 0.9cm 1cm 1cm 1cm,scale=0.8]{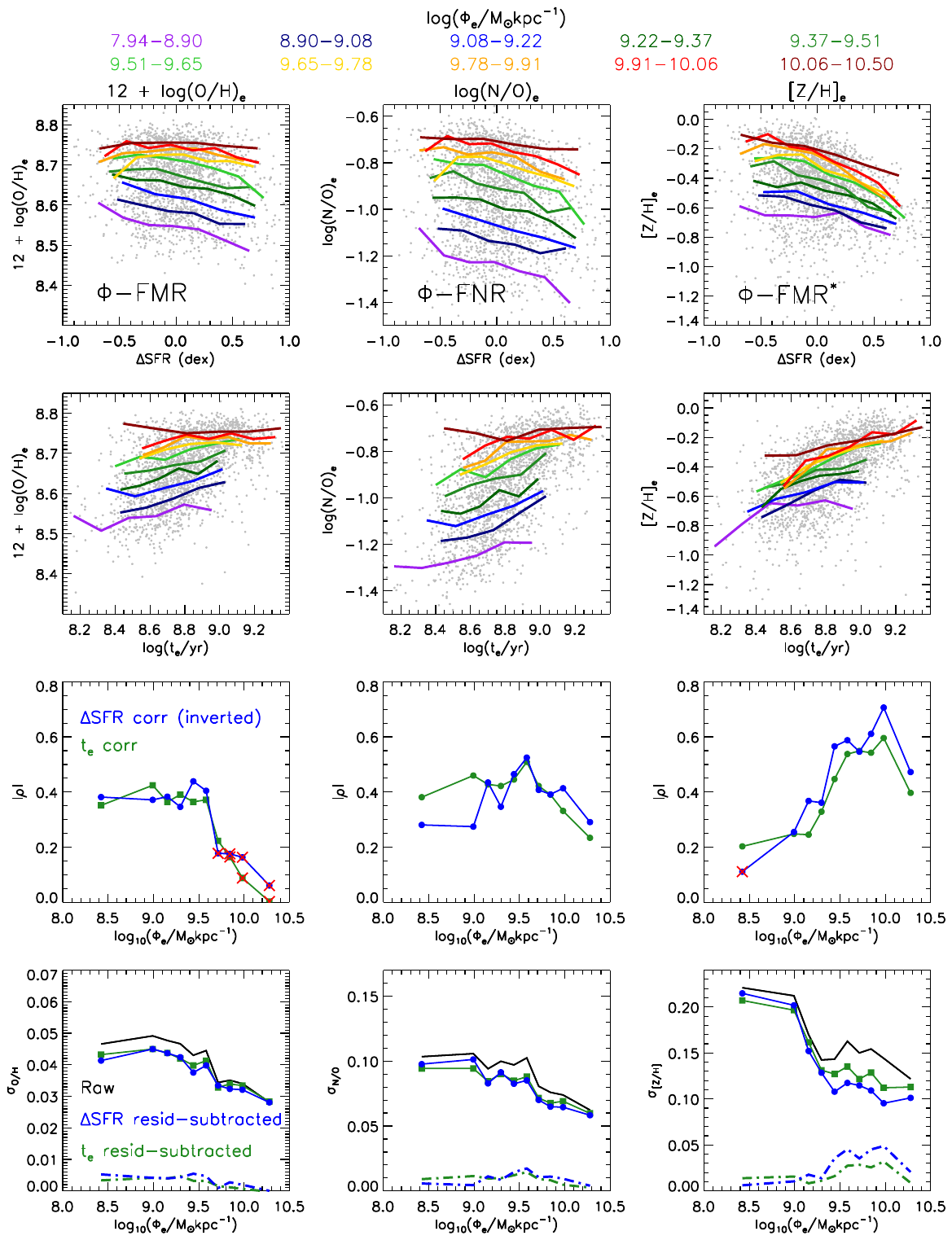} 
	\caption{As in \autoref{fmr_assess}, but for bins of gravitational potential each comprising $\sim$10 \%\ of sample galaxies. We show results for the $\Phi$-FMR (left), $\Phi$-FNR (middle) and $\Phi$-FMR$^*$ (right) in turn, along with equivalent relations involving $t_e$.}
	\label{potfmr_assess}
	\end{center}
\end{figure*}

\begin{figure*}
\begin{center}
	\includegraphics[trim = 0.5cm 1cm 3cm 11cm,scale=0.8]{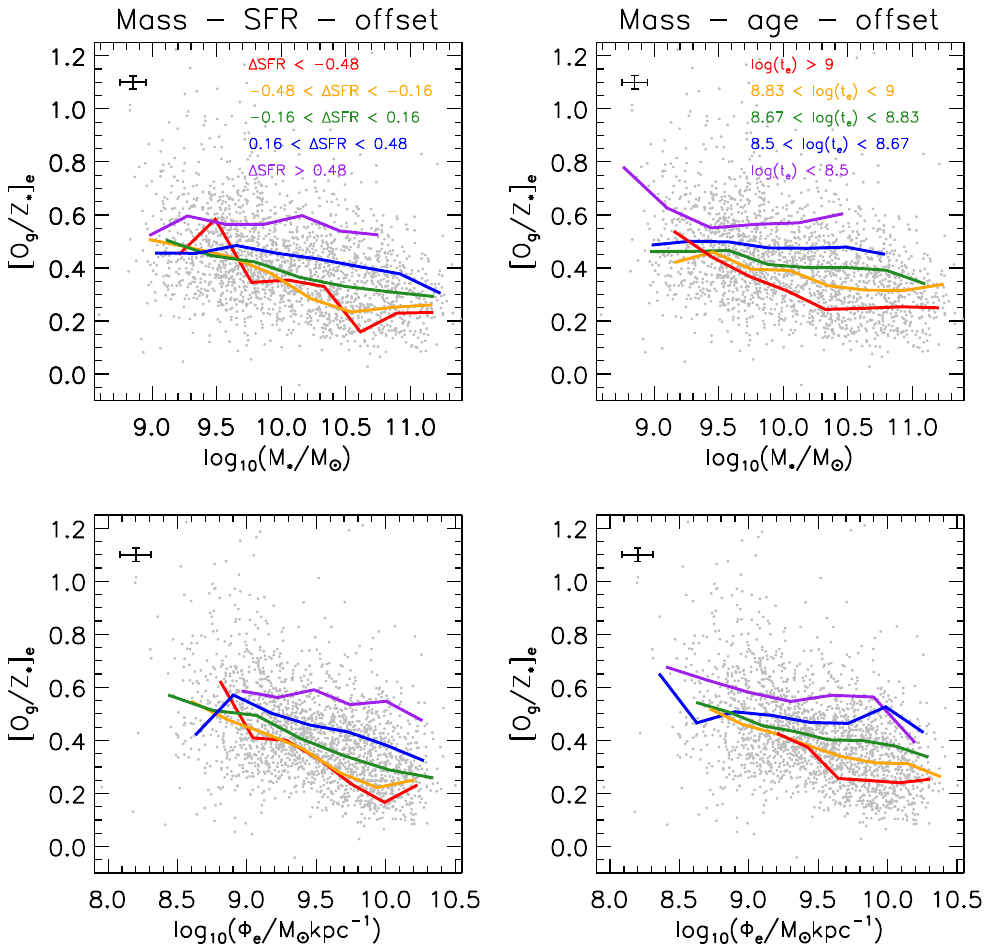} 
	\caption{Gas-stellar metallicity offset $\mathrm{[O_g/Z_*]_e}$ at 1 $R_e$ as a function of $M_*$ (top) and $\mathit{\Phi_e}$ (bottom). The colored lines show medians in bins of $M_*$ (top) or $\mathit{\Phi_e}$ (bottom) for different $\mathrm{\Delta SFR}$ ranges. We detect an inverse correlation between $\mathrm{[O_g/Z_*]_e}$ and $\mathrm{\Delta SFR}$ across almost the full range in $M_*$ and $\mathit{\Phi_e}$. The black error bars show median errors.}
	\label{fmr_offset}
	\end{center}
\end{figure*}

From \autoref{fmr_assess}, we find the FMR to be weak as in previous IFU analyses \citep{sanchez2013, sanchez2017a, bb2017}: it yields statistically insignificant residual trends ($\mathrm{P > 0.01}$ in half of the mass bins, with the weakest trends seen in the two highest-mass bins. The FMR is, however, \rev{stronger than the FNR or FMR$^*$} in the lowest-mass bin. By contrast, the FNR and FMR$^*$ are stronger at higher masses, with the strongest trends --- based on both $\rho$ and on $\sigma$ reductions --- found at masses $\mathrm{\sim 10^{10.5} \mathrm{M_\odot}}$. Our findings support a link between abundances and SFRs over the full mass range of our sample, contrary to what would be concluded from the FMR alone. 

We also find from \autoref{fmr_assess} that $t_e$ is a more constraining third parameter than SFR when the FMR and FNR are considered. This is apparent visually from the figure's top two rows, and it can also be seen from the increase in $\rho$ and reduction in $\sigma$ over most mass bins.

In \autoref{potfmr_assess}, we present the $\Phi$-FMR, $\Phi$-FNR and $\Phi$-FMR$^*$ in the same format as for \autoref{fmr_assess}, along with presenting equivalent relations in which $t_e$ replaces $\mathrm{\Delta SFR}$. Our use of $\mathit{\Phi_e}$ is motivated by the strong observed correlations between $\mathit{\Phi_e}$ and the abundance measures we consider, along with the clear residual trends at a given $\mathit{\Phi_e}$ evident in Figures 3 and 4.

From comparing \autoref{potfmr_assess} to \autoref{fmr_assess}, we make four key observations regarding gaseous abundances:

\begin{itemize}
\item From the two figures' bottom rows, the $\Phi$ZR and $\Phi$NR possess significantly lower dispersions than do the FMR and FNR at nearly all masses and $\mathit{\Phi_e}$ values. \rev{The FMR's dispersion for instance exceeds 0.06 dex in a couple mass bins, while the $\Phi$ZR dispersion is below 0.05 dex for all $\Phi_e$ bins. The FNR's dispersion peaks at above 0.14 dex, meanwhile, while the $\Phi$NR is always below 0.11 dex. Thus,} \textit{$\mathit{\Phi_e}$ is more informative than $M_*$ and SFR combined, for understanding gas abundances.}
\item From $\rho$ and $\sigma$, gas abundances trend more tightly with $\mathrm{\Delta SFR}$ at a given $\mathit{\Phi_e}$ than at a given $M_*$ in almost all cases. The combination of $\mathit{\Phi_e}$ and $\mathrm{\Delta SFR}$ thus captures significantly more abundance scatter than when we combine $\mathrm{\Delta SFR}$ with $M_*$. \textit{The FMR and FNR not only persist, but significantly strengthen, if we replace $M_*$ with $\mathit{\Phi_e}$. Gaseous chemical abundances are sensitive to galaxies' star-formation states, even after taking $\mathit{\Phi_e}$ (i.e., stellar mass and size) into account}.
\item When starting from $\mathit{\Phi_e}$-abundance relations, $\mathrm{\Delta SFR}$ and $t_e$ perform very similarly as third parameters in terms of $\rho$ and $\sigma$. \textit{SFR and stellar age are roughly as informative as each other on chemical abundances}, and are possibly providing overlapping information on galaxies' SFHs.
\item The $\Phi$-FMR persists over a significant $\mathit{\Phi_e}$ range but still disappears at higher $\mathit{\Phi_e}$ values, while the $\Phi$-FNR persists over the full tested $\mathit{\Phi_e}$ range.

\end{itemize}

From considering $\rho$ and $\sigma$ from stellar metallicity relations (right-hand columns in Figures 5 and 6), we find stronger residual trends at high $M_*$ or high $\mathit{\Phi_e}$ than were obtained for the gaseous abundances. We further find that the FMR$^*$ and $\Phi$-FMR$^*$ are roughly as strong as each other, again from considering $\rho$ and $\sigma$; \textit{replacing $M_*$ with $\mathit{\Phi_e}$ has almost no impact on the strength of recovered $\mathit{[Z/H]_e}$ relations}. Though $\mathit{\Phi_e}$ is slightly more informative then $M_*$ of the stellar metallicity (e.g. \autoref{age_colorre}, top panels), \textit{most added information on $\mathit{[Z/H]_e}$ is instead gained by considering the star-formation state of a galaxy (as parametrised by $\mathit{\Delta SFR}$ or $\mathit{t_e}$)}.

\begin{figure*}
\begin{center}
	\includegraphics[trim = 0cm 7cm 6cm 9cm,scale=1]{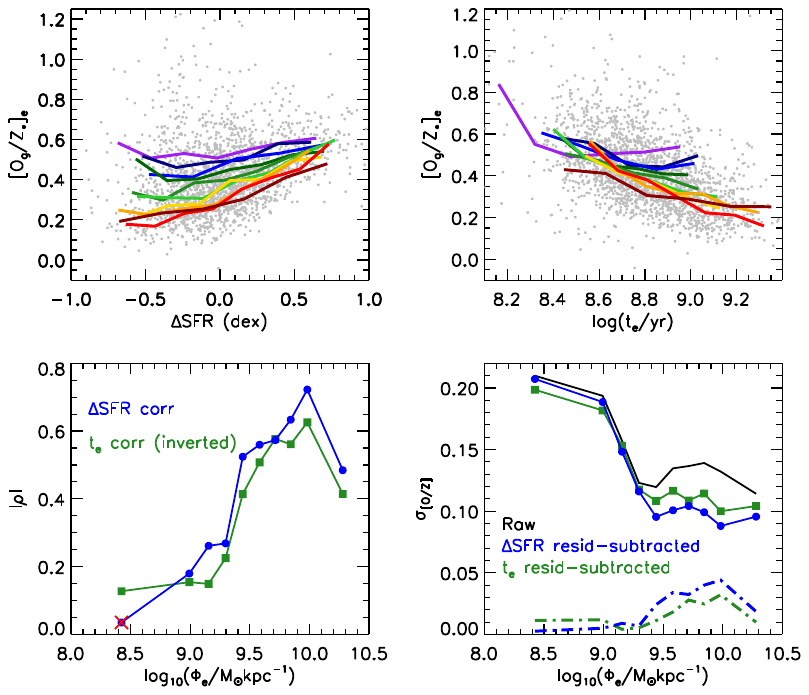} 
	\caption{Top left: gas-stellar metallicity offset $\mathrm{[O_g/Z_*]_e}$ as a function of $\mathrm{\Delta SFR}$, in bins of $\mathit{\Phi_e}$; line colours are as in \autoref{potfmr_assess}. Top right: $\mathrm{[O_g/Z_*]_e}$ as a function of stellar age, in bins of $\mathit{\Phi_e}$. Bottom left: Spearman correlation coefficients between $\mathrm{\Delta SFR}$ (or $t_e$) and $\mathrm{[O_g/Z_*]_e}$, in bins of $\mathit{\Phi_e}$; we again show absolute values of the coefficients for easier comparison, with red Xs indicating statistically insignificant ($P > 0.01$) correlations. Bottom right: dispersion in $\mathrm{[O_g/Z_*]_e}$ before (black lines) and after subtracting out fitted $\mathrm{\Delta SFR}$--abundance relations (blue circles) or fitted age--abundance relations (green squares); the colored dot-dashed lines at the bottom indicate the achieved reductions in scatter.}
	\label{potfmroffsets_assess}
	\end{center}
\end{figure*}

\subsection{Assessing the stellar-gas metallicity offset}\label{sec34}

We next consider the ratio of gaseous and stellar metallicities, $\mathrm{[O_g/Z_*]_e}$; this is equal to $\mathrm{12 + \log(O/H)_e - 8.69 - [Z/H]_e}$, where 8.69 is the assumed solar oxygen abundance \citep{asplund2009}. We show in \autoref{fmr_offset} that $\mathrm{[O_g/Z_*]_e}$ positively correlates with $\mathrm{\Delta SFR}$ and negatively correlates with $t_e$ at a given $M_*$ or $\mathit{\Phi_e}$. We also find that $\mathrm{[O_g/Z_*]_e}$ inversely correlates with $M_*$ and $\mathit{\Phi_e}$, similarly to what \citet{frasermckelvie2022} report from SAMI data. 

Starting from the $\mathit{\Phi_e}$--$\mathrm{[O_g/Z_*]_e}$ relation, we assess in \autoref{potfmroffsets_assess} the residual $\mathrm{[O_g/Z_*]_e}$ trends with $\mathrm{\Delta SFR}$ and with $t_e$ in turn. As before, we carry out our assessment using $\rho$ and $\sigma$. We find significant relations in this case, with $[O_g/Z_*]$ displaying significant negative (positive) correlations with $\mathrm{\Delta SFR}$ ($t_e$) at high $\mathit{\Phi_e}$. Given the results from \autoref{potfmr_assess} presented in the previous subsection, the behaviour of $\mathrm{[O_g/Z_*]_e}$ appears to be driven largely by $\mathrm{[Z/H]_e}$: the $\Phi$-FMR is at its weakest at high $\Phi_e$ values, while the $\Phi$-FMR$^*$ strengthens significantly at high $\Phi_e$.

\section{Discussion}\label{disc}

We have used MaNGA data to assess how stellar and gaseous chemical abundances vary with SFR and stellar age at a given value of $M_*$ or $M_*/R_e$ ($\mathit{\Phi_e}$), with the latter frequently treated as a proxy for gravitational potential. Our key results can be summarised as follows:

\begin{itemize}
\item Star-forming galaxies show a connection between star-formation state and chemistry at the full range of tested $M_*$ and $\mathit{\Phi_e}$ values. 
\item $\mathit{\Phi_e}$ is significantly more predictive of gas abundances than is $M_*$ and SFR together, making $\mathit{\Phi_e}$ a much better starting point for understanding the physics behind gaseous abundances in galaxies than the more commonly explored FMR. 
\item At higher values of $M_*$ the traditional FMR vanishes in agreement with previous work; the $M_*$--SFR--N/O relation (FNR) persists at higher masses, meanwhile, with the stellar FMR$^*$ strengthening significantly.
\item Equivalent results to the previous point are obtained when considering $\mathit{\Phi_e}$ instead of $M_*$, with the resulting relations being tighter for all three abundance measures.

\end{itemize}

\begin{figure*}
\begin{center}
	\includegraphics[trim = 3cm 8cm 0cm 8cm,scale=0.75]{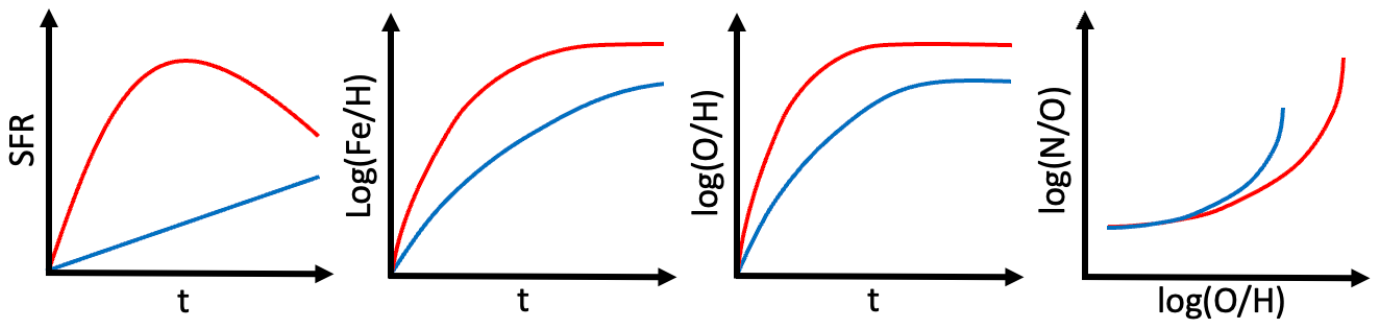} 
	\caption{Cartoon depiction of two example SFH shapes plotted as SFR vs time (leftmost panel), and their resulting chemical evolution histories in terms of Fe/H (left middle), O/H (right middle), and in terms of their N/O--O/H relations (rightmost panel). We assume the early SFH (red lines) to possess a higher $\mathit{\Phi_e}$ than the later SFH (blue lines). All three abundances represent ISM values at a particular time.}
	\label{sfh_cartoon}
	\end{center}
\end{figure*}

We provide a physical interpretation of our results in \autoref{disc_interp1} and \autoref{disc_interp2} . We then discuss the implications of our findings for observations (\autoref{disc_obs}) and simulations (\autoref{disc_sims}) separately. 

\subsection{A unified interpretation of the $\Phi$ZR, $\Phi$NR and $\Phi$ZR$^*$}\label{disc_interp1}

To understand our findings, we begin with the $\Phi$ZR and $\Phi$NR. These were shown in \citet{boardman2024a} to both be tight (Spearman's $\rho > 0.8$) in MaNGA, with the $\Phi$ZR appearing to be driven by a combination of the $\Phi$NR and the N/O--O/H relation. $\mathit{\Phi_e}$ encodes information on a wide range of relevant galaxy properties, with higher-$\mathit{\Phi_e}$ galaxies on average being older in addition to being more massive and possessing higher escape velocities. 

The importance of $\mathit{\Phi_e}$ is often ascribed to varying outflow efficiencies as set by the escape velocity \citep[e.g.][]{deugenio2018}. However, there are a number of challenges to this interpretation. Gas metallicities in the EAGLE simulations show almost no dependence on escape velocity at a given $M_*$ \citep{sm2018a}, while observational gas metallicities show relatively weak trends with galaxy dynamical masses \citep{baker2023}. It must also be remembered that the gas-stellar metallicity offset negatively correlates with both $M_*$ and $\mathit{\Phi_e}$, which cannot be straightforwardly explained by variations in escape velocity alone; this can be seen in our \rev{Figures 7 and 8}, and it has also been reported from SAMI data by \citet{frasermckelvie2022}. That the $\Phi$NR appears to drive the $\Phi$ZR \citep{boardman2024a}, and not vice-versa, further challenges escape velocities as the primary factor driving the $\mathit{\Phi_e}$ parameter's importance.

An alternative possibility is that $\mathit{\Phi_e}$ is so predictive of chemical abundances due to the connection of $\mathit{\Phi_e}$ to a galaxy's SFH \citep[e.g.][]{barone2018,barone2020}, with more compact galaxies expected to experience earlier and more rapid SFHs. More rapid SFHs result in larger proportions of gas reservoirs being consumed and converted into metals, resulting in higher gas metallicities and producing a tight $\Phi$ZR at low redshifts. A stellar $\Phi$ZR$^*$ then follows from the gaseous $\Phi$ZR, due to stellar metallicity effectively representing a weighted average of the gas metallciity. The stellar relation possesses a larger scatter than the gaseous relation; \rev{this can be attributed in part to variations in a galaxy's metallicity history which in turn can be attributed to SFH variations at earlier times, to which gaseous abundances will be less sensitive, \rev{though increased uncertainties in stellar metallicities (when compared to gaseous metallicities)} may also contribute}. 

The $\Phi$ZR in turn produces a tight $\Phi$NR. This occurs due to the nitrogen enrichment rate rising with metallicity, which also produces the tight N/O--O/H relation observed at higher metallicities \citep[e.g.][]{andrews2013}. That the $\Phi$NR is even tighter than the $\Phi$ZR in MaNGA \citep{boardman2024a} can be ascribed to the increased sensitivity of N/O to the shape of a galaxy's SFH (as encoded by $\mathit{\Phi_e}$), along with O/H being more susceptible to being scattered by recent metal-poor inflows. 

\begin{figure*}
\begin{center}
	\includegraphics[trim = 0cm 3cm 0cm 3cm,scale=0.59]{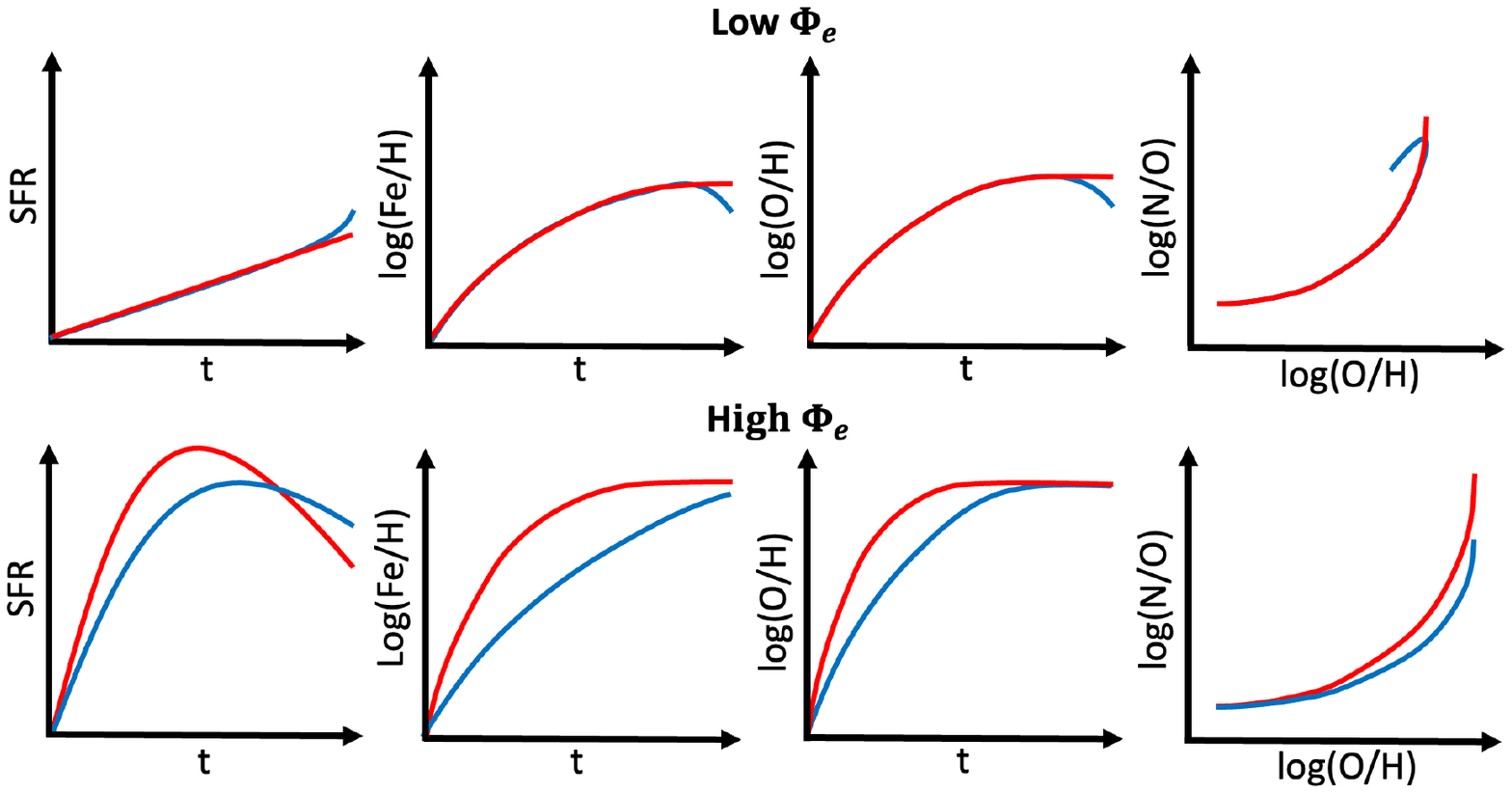} 
	\caption{Cartoon depiction of four more example SFH shapes plotted as SFR vs time (leftmost panels), and their resulting chemical evolution histories in terms of Fe/H (left middle panels), O/H (right middle pabels) and in terms of their N/O--O/H relations (rightmost panels). All three abundances represent ISM values. In the top panels, the blue and red lines depict a pair of low-$\mathit{\Phi_e}$ galaxies with diverging recent inflow histories; this results in differences in the galaxies' \rev{ISM Fe/H, O/H and N/O abundances} while having little impact on the measured stellar metallicity, which represents a weighted average of the metallicity history. In the bottom panels, the red and blue lines instead correspond to two high-$\mathit{\Phi_e}$ with slight differences in their broad SFH shape; the two galaxies have equal O/H abundances in this case, but the different chemical evolution histories produce different gaseous N/O abundances and different stellar metallicities.}
	\label{sfh_cartoon_hilow}
	\end{center}
\end{figure*}

Building on the arguments of \citet{frasermckelvie2022}, we present in \autoref{sfh_cartoon} two example SFH shapes --- an early SFH and a more gradual late SFH --- and the resulting chemical evolution histories in terms of O/H, Fe/H and N/O. If we now assume the earlier SFH to correspond to a higher $\mathit{\Phi_e}$ value and hence to a higher gas metallicity, then we conclude that the $\Phi$ZR, $\Phi$NR and $\Phi$ZR$^*$ all arise from earlier and/or more rapid SFHs in deep potential wells. The plotted Fe/H here is effectively an \textit{ISM} Fe/H, which does not consider depletion onto dust; the measured stellar [Z/H], which is mostly sensitive to the stellar [Fe/H], instead represents a weighted average of the ISM Fe/H over a galaxy's lifetime. We assume iron to enrich more slowly than oxygen, due to the longer timescales for iron production in stars \cite[e.g.][]{maiolino2019}.

Our depicted N/O--O/H evolution in \autoref{sfh_cartoon} is informed by \citet{johnson2023}, who study nitrogen and oxygen enrichment using the Milky Way chemical evolution model presented in \citet{johnson2021}. In this model, oxygen initially enriches at near-constant N/O due to core-collapse supernovae, with N/O then rising at a metallicity-dependent rate due to nitrogen enrichment from AGB stars. At late times in this model, N/O continues to rise while O/H remains near-constant. Enrichment therefore does \textit{not} proceed along the shape of the measured N/O--O/H relation at low redshifts \citep[e.g.][]{pilyugin2010,andrews2013,Dopita_2016_EmLineDiagnostic,pm2016}, contrary to what is often assumed \citep[e.g.][]{koppen2005,molla2006,vincenzo2016}. 

In \autoref{sfh_cartoon}, the earlier SFH produces a faster rate of metal-enrichment, with oxygen and iron  also enriching to higher equilibrium values than for the late SFH; this results in the early SFH yielding a higher measured gas metallicity and a higher measured stellar metallicity. The final gaseous N/O is also higher for the earlier SFH, due to a greater portion of old stars and also due to stars' higher metallicities.

\subsection{A unified interpretation of the FMR, FNR and FMR$^*$}\label{disc_interp2}

At a given $\mathit{\Phi_e}$, variations in $\mathrm{\Delta SFR}$ can be interpreted as reflecting SFH variations. If more stars form early, then more gas is consumed early, resulting in lower $\mathrm{\Delta SFR}$ at late times. If significant quantities of gas are accreted recently, then this produces more star-formation, higher $\mathrm{\Delta SFR}$ and lower measured stellar ages. As such, $\mathrm{\Delta SFR}$ and $t_e$ provide similar information, with their connection to chemical abundances reflecting the importance of galaxy SFHs.

At the lowest $\mathit{\Phi_e}$ values, variations in $\mathrm{\Delta SFR}$ or $t_e$ produce variations in gas abundances while having little impact on $\mathrm{[Z/H]}$. For these low $\mathit{\Phi_e}$ values, we argue that the FMR and FNR result from variations in galaxies' recent inflow rates leading to variations in recent SFHs. We further argue that these variations occur on timescales too short for stellar populations' $\mathrm{[Fe/H]}$, and hence the measured $\mathrm{[Z/H]}$, to be significantly affected. That gaseous N/O is affected could imply that the accreting gas is metal-poor but not pristine \citep[e.g.][]{hp2022}, or else that the variation timescales are long enough for oxygen enrichment to have occurred after pristine inflows \citep{boardman2024}.

As we look to higher $\mathit{\Phi_e}$ values, we see that variations in $\mathrm{\Delta SFR}$ (or $t_e$) become increasingly associated with $\mathrm{[Z/H]}$ variations, with gaseous O/H variations largely disappearing. In this case, we argue that variations in $\mathrm{\Delta SFR}$ or $t_e$ relate to longer-term SFH variations rather than to recent changes. Earlier SFHs will correspond to more rapid initial enrichment and greater consumption of galaxies' gas reservoirs, resulting in higher measured stellar metallicities. Earlier SFHs will also result in older stellar populations, producing greater nitrogen enrichment at a given gas metallicity. The measured gaseous O/H is largely unaffected, meanwhile, because it is sensitive almost entirely to the \textit{present} properties of a given galaxy.

In \autoref{sfh_cartoon_hilow}, we present two further sets of cartoons. In the top set, we show a pair of low-$\mathit{\Phi_e}$ galaxies with slightly different recent inflow histories, leading their SFHs to diverge at late times. The galaxy with a higher present SFR is receiving infalling gas at greater rate by the time of observation, which lowers the gaseous O/H and Fe/H. The gaseous N/O is also reduced, either due to the incoming gas being non-pristine with low N/O or else due to the gas quickly enriching in oxygen via core-collapse supernovae. The measured stellar metallicities are barely affected however, due to the vast majority of stars having formed before the inflow.

In the bottom set of \autoref{sfh_cartoon_hilow} cartoons, we show a pair of high-$\mathit{\Phi_e}$ galaxies with slightly different SFH shapes, which again produce variations in recent star-formation activity. In this case, the SFH variations are longer-term, and they produce differences in the chemical evolution histories without affecting the final oxygen abundances. The earlier SFH produces a higher rate of nitrogen enrichment at fixed O/H, however, and so it raises the measured N/O. The measured stellar metallicity is also significantly higher for the earlier SFH, due to iron enrichment proceeding much faster. 

\subsection{Implications: observations}\label{disc_obs}

Our results demonstrate a connection between star-formation properties and abundances over the full tested range of $M_*$ and $\mathit{\Phi_e}$ ($=M_*/R_e$). Thus, $M_*$, size, and star-formation state are all relevant in understanding the chemistries of galaxies. This conclusion was reached by considering gaseous N/O and stellar metallicity in addition to the gas metallicity O/H, which on its own shows a diminishing connection with star-formation state at higher values of $M_*$ or $\mathit{\Phi_e}$. The traditional FMR in fact appears to be among the weakest of the relations considered in this work, except at the lowest tested stellar masses. Replacing $M_*$ with $\mathit{\Phi_e}$ produces a notably tighter relation than the FMR, as does replacing SFR with stellar age. 

Our results for stellar metallicities are in good agreement with \citet{looser2024} which in turn is at odds with \citet{vaughan2022}, who find essentially no difference in the stellar $\Phi$ZRs for star-forming and passive galaxies; \citet{vaughan2022} thus argue that galaxy size can explain metallicity offsets between star-forming and passive galaxies. Such a conclusion is not supported by MaNGA, in which a clear three-way relation exists between potential, star-formation rate and stellar metallicity, both from \citet{looser2024} and independently from this present work. \citet{looser2024} draw their conclusions from metallicities within 1~$R_e$, which is effectively intermediate between our work (which uses metallicities \textit{at} 1~$R_e$) and \citet{vaughan2022} (which uses central metallicities). Given our agreement with \citet{looser2024}, disagreements between MaNGA and SAMI are likely due to sample selection differences rather then due to differences in methodology. The MaNGA sample contains more high-$M_*$ galaxies than the SAMI sample, as pointed out by \citet{looser2024}, which possibly makes MaNGA better-suited for studying metallicity trends in high-$\mathit{\Phi_e}$ star-forming galaxies.     

Our results add to a growing body of literature concerning the importance of size in setting galaxy properties. Even for star-forming galaxies specifically, size --- parametrised by half-light radius or else by similar measures such as radial disc scale lengths \citep[e.g.][]{simard2011} --- has been shown to possess relations with numerous other properties at a given stellar mass. These properties include SFRs \citep[e.g.][]{wuyts2011}, bulge-to-total ratios \citep[e.g.][]{boardman2022}, stellar population ages \citep[e.g.][]{barone2020}, gaseous and stellar chemical abundances \citep[e.g.][]{ellison2008,barone2020} and radial gas abundance gradients \citep{boardman2021}. Local gas abundances (on $\sim$kpc scales) also appear to trend with other local properties differently in compact galaxies compared to in extended galaxies, from MaNGA observations \citep{schaefer2022,boardman2022,boardman2023}.

For various extragalactic science questions, it can be desirable to match a specific sample of galaxies to a so-called control sample. Such questions include ones relating to quenching and SFRs \citep[e.g.][]{frazermckelvie2018,sharma2023}, the effects of galaxies' enviornments \citep[e.g.][]{fernandez2024}, the effects of galaxy interactions \citep[e.g.][]{perez2009,alonso2012}, the impact of galaxy morphological features \citep[e.g.][]{vera2016}, the impact of starbursts \citep[e.g.][]{setton2022}, star-gas misalignments \citep[e.g.][]{zhou2022}, and many more besides. In such cases, the control sample is typically selected on a combination of stellar mass and redshift, with morphological information also sometimes used \citep[e.g.][]{fernandez2024}. Stellar mass is reasoned to be a particularly informative property of galaxies, with redshift accounting for the effects of cosmic time \citep[e.g.][]{ellison2008a,perez2009}. For many science questions however, \textit{it will be appropriate to select on size in addition to other parameters}. The variation in galaxies' properties with size is clearly significant at fixed mass. Therefore, by selecting control samples also on size, the resulting control samples will be significantly more meaningful.

Current and upcoming spectroscopic surveys --- including those from DESI, PFS  \citep{greene2022}, MOONS \citep{cirasuolo2020,maiolino2020} and 4MOST \citep{dejong2019,walcher2019} --- will significantly increase the number of galaxy spectra observed at high redshifts ($\mathrm{z} \gtrsim 1$). As mentioned in Section 1, the $\Phi$ZR should evolve significantly with redshift. This evolution should be larger than the substantial redshift evolution observed in the MZR, given that higher-redshift galaxies are both metal-poorer \citep[e.g.][]{maiolino2008,zahid2014,jones2020} and more compact \citep{vdw2014} at a given mass. Given the tightness of the $\Phi$ZR at low redshifts, it would be highly informative to track the evolution of this relation with redshift, as well as to track the redshift evolution of the $\Phi$-FMR. It would likewise be informative to track the redshift evolution of the $\Phi$-FMR$^*$ and $\Phi$-FNR, given the observed abundance--SFR connection over the full $\mathit{\Phi_e}$ range in this work. Finally, high-redshift spectroscopy will let us track how the star-forming  galaxy population evolves in N/O--O/H space, which in turn will help test the evolutions proposed in Figures 9 and 10 (right panels) for individual galaxies. 

\subsection{Implications: simulations}\label{disc_sims}

At a given stellar mass in observations, gas metallicity correlates negatively with both size measures (e.g. $\mathrm{R_e}$) and with SFR. Various large scale cosmological hydrodynamical simulations are able to reproduce this trend, including EAGLE and IllustrisTNG \citep{sm2018a,ma2024}. Both of these simulations display a positive size-SFR correlation at fixed stellar mass over a wide mass range \citep{sm2018a,ma2024}. The residual size--metallicity and SFR--metallicity trends can therefore be ascribed to the same source in simulations, with \citet{sm2018a} ascribing them to long-term inflow variations and hence to long-term SFH variations.

However, for star-forming galaxies at fixed stellar mass over a significant mass range, \textit{size and SFR do not positively correlate in observations}. This can for instance be seen in \citet{wuyts2011} (their fig 3) as well for MaNGA in our \autoref{sampleplot} (middle and bottom panels) and in \citet{jf2024}. We verified in \autoref{sampleplot} that half-light and half-mass radii behave similarly in our sample; this means that switching to a mass-based size measure would not resolve the tension with simulations, leaving it unclear whether the simulation-based view can be applied to observed galaxies.

In order to simultaneously probe size--metallicity and SFR--metallicity trends at fixed mass (as well as to probe equivalent relations with gas metallicity or with other abundance measures), it is therefore vital that simulations first reproduce galaxies' size--SFR behaviour accurately. After this, one should confirm the simultaneous recovery of the FMR and $\Phi$ZR in simulations. The $\Phi$ZR is significantly tighter than the FMR in observations \citep[see also][]{ma2024a}, which is another aspect that should ideally be reproduced in simulations. Simulations could then be used to probe the interpretation provided in \autoref{sec31} and \autoref{sec32} above, in which chemical abundances are shaped by a combination of short-term gas inflow histories and longer-term star-formation histories, as well as to seek alternate interpretations if required. 

\section{Summary and Conclusion}

Gaseous chemical abundances represent the end products of galaxy evolution, while stellar chemical abundances represent a weighted integral of a galaxy's chemical evolution. Thus, we can learn much about galaxy chemical evolution by considering gaseous and stellar abundances together. To this end, we studied stellar and gaseous abundances at 1 effective radius ($R_e$) in star-forming galaxies, using data from the SDSS-IV MaNGA survey. 

For chemical abundance measures, we focused on gaseous O/H and N/O abundances along with light-weighted stellar metallicity, which is primarily sensitive to the stellar iron abundance in galaxies. We then considered residual trends in chemical abundances at a given $M_*$ or  $\mathit{\Phi_e}$ ($ = M_*/R_e$), in terms of SFR or light-weighted stellar age ($t_e$). $\mathit{\Phi_e}$ ($ = M_*/R_e$) has previously been shown to be an especially tight predictor of galaxies' chemical abundances \citep[e.g.][]{deugenio2018,barone2018,barone2020,boardman2024a}, with residual trends from it largely unexplored until now. We focused on residual trends involving SFR or $t_e$) due to these both being representative of galaxies' star-formation states.

We confirm from our results the gaseous and stellar fundamental metallicity relations (FMR and FMR$^*$, between stellar mass, SFR and metallicity) in MaNGA, along with detecting the fundamental nitrogen relation (FNR, between stellar mass, SFR and the gasous N/O abundance). Though the gaseous FMR disappears at higher mass, the FNR persists and the stellar FMR$^*$ significantly strengthens. We further noted the gaseous relations to strengthen when replacing SFR with stellar age. From our results, star-forming galaxies show a connection between chemical abundances \rev{and} star-formation state at any given stellar mass within our sample's mass range, which is a conclusion which would not be reached from studying the gaseous FMR in isolation. 

We found that replacing stellar mass with $\mathit{\Phi_e}$ tightens the three `fundamental' relations, particularly the gaseous FMR and FNR; we termed these modified relations the $\Phi$-FMR, $\Phi$-FNR and $\Phi$-FMR$^*$. We further found that $\mathit{\Phi_e}$ is more predictive of gaseous O/H and N/O \textit{on its own} than $M_*$ and SFR are \textit{together}. We therefore argue that $\mathit{\Phi_e}$ is a much better starting point than $M_*$ for understanding the physics underlying gaseous abundances. Galaxies continue to show a connection between abundances and star-formation state at any given $\mathit{\Phi_e}$, suggesting that both $\mathit{\Phi_e}$ and star-formation state are relevant for understanding abundance in galaxies. Similarly to the stellar mass case, the $\Phi$-FMR disappears at higher values of $\mathit{\Phi_e}$ while the $\Phi$-FNR persists and the $\Phi$-FMR$^*$ strengthens.

From these results, we presented a unified interpretation of gaseous O/H, gaseous N/O and stellar metallicity in star-forming galaxies. \textit{Higher $\mathit{\Phi_e}$ is associated with higher present-day O/H due to being associated with earlier SFHs}, producing the observed tight $\Phi$ZR. This naturally produces a tight $\Phi$NR, due to the nitrogen enrichment rate depending on metallicity, along with producing a tight $\Phi$ZR$^*$. \textit{At low $\mathit{\Phi_e}$, chemical abundances vary with SFR (or with stellar age) due to recent inflow-induced SFH variations. At higher $\mathit{\Phi_e}$, abundances instead vary due to longer-term SFH variations that correspond to large-scale variations in gas consumption histories.}

The above interpretation does not directly consider post-starburst (PSB) galaxies. These are galaxies that have experienced strong starbursts followed by rapid star-formation cessation, producing an excess of $\sim$ Gyr-old stars \citep[e.g.][]{zabludoff1996}. PSBs are notably compact for their mass \citep[e.g.][]{almaini2017,chen2022}, and they appear to enrich substantially following the starburst in most cases \citep{leung2024}. PSB galaxies are rare in the nearby Universe but are much more common at higher redshifts \citep[e.g.][]{wild2020}. We will investigate chemical abundances of PSB galaxies for future work, using chemical evolution models. 

Our results have implications both for simulations and for future observations. It is crucial that simulations correctly recover galaxies' size--SFR behavior at a given mass, prior to simulations being used to investigate size--metallicity or SFR--metallicity behavior. Tracking the evolution of $\mathit{\Phi_e}$-based abundance relations and the N/O--O/H relation in star-forming galaxies over cosmic time will be possible with upcoming galaxy surveys, and will further confirm our unified interpretation.

\section*{Acknowledgements}

NFB and VW acknowledge Science and Technologies Facilities Council (STFC) grant ST/Y00275X/1. NVA acknowledges support of Conselho Nacional de Desenvolvimento Cient\'{i}fico e Tecnol\'{o}gico (CNPq). FDE acknowledges support by the Science and Technology Facilities Council (STFC), by the ERC through Advanced Grant 695671 ``QUENCH'', and by the UKRI Frontier Research grant RISEandFALL. We thank Rita Tojeiro for providing a number of useful comments on an early version of this manuscript, which helped significantly towards this manuscript's development. Funding for the Sloan Digital Sky Survey IV has been provided by the Alfred P. Sloan Foundation, the U.S. Department of Energy Office of Science, and the Participating Institutions. SDSS-IV acknowledges support and resources from the Center for High-Performance Computing at the University of Utah. The SDSS web site is \url{www.sdss4.org}. 

SDSS-IV is managed by the Astrophysical Research Consortium for the Participating Institutions of the SDSS Collaboration including the Brazilian Participation Group, the Carnegie Institution for Science, Carnegie Mellon University, the Chilean Participation Group, the French Participation Group, Harvard-Smithsonian Center for Astrophysics, Instituto de Astrof\'isica de Canarias, The Johns Hopkins University, Kavli Institute for the Physics and Mathematics of the Universe (IPMU) / University of Tokyo, Lawrence Berkeley National Laboratory, Leibniz Institut f\"ur Astrophysik Potsdam (AIP),  Max-Planck-Institut f\"ur Astronomie (MPIA Heidelberg), Max-Planck-Institut f\"ur Astrophysik (MPA Garching), Max-Planck-Institut f\"ur Extraterrestrische Physik (MPE), National Astronomical Observatories of China, New Mexico State University, New York University, University of Notre Dame, Observat\'ario Nacional / MCTI, The Ohio State University, Pennsylvania State University, Shanghai Astronomical Observatory, United Kingdom Participation Group, Universidad Nacional Aut\'onoma de M\'exico, University of Arizona, University of Colorado Boulder, University of Oxford, University of Portsmouth, University of Utah, University of Virginia, University of Washington, University of Wisconsin, Vanderbilt University, and Yale University.

\section*{Data Availability}

All data used here are publically available. MaNGA data and analysis products can be accessed via the Marvin interface \citep{cherinka2019}, both online\footnote{\url{https://magrathea.sdss.org/marvin/}} and with \textsc{PYTHON}\footnote{\url{https://sdss-marvin.readthedocs.io/en/latest/installation.html}}. MaNGA data and products can also be downloaded from the SDSS Science Archive server\footnote{\url{https://data.sdss.org/sas/}}.

\bibliographystyle{mnras}
\bibliography{bibliography}

\begin{thebibliography}{}
\makeatletter
\relax
\def\mn@urlcharsother{\let\do\@makeother \do\$\do\&\do\#\do\^\do\_\do\%\do\~}
\def\mn@doi{\begingroup\mn@urlcharsother \@ifnextchar [ {\mn@doi@} {\mn@doi@[]}}
\def\mn@doi@[#1]#2{\def\@tempa{#1}\ifx\@tempa\@empty \href {http://dx.doi.org/#2} {doi:#2}\else \href {http://dx.doi.org/#2} {#1}\fi \endgroup}
\def\mn@eprint#1#2{\mn@eprint@#1:#2::\@nil}
\def\mn@eprint@arXiv#1{\href {http://arxiv.org/abs/#1} {{\tt arXiv:#1}}}
\def\mn@eprint@dblp#1{\href {http://dblp.uni-trier.de/rec/bibtex/#1.xml} {dblp:#1}}
\def\mn@eprint@#1:#2:#3:#4\@nil{\def\@tempa {#1}\def\@tempb {#2}\def\@tempc {#3}\ifx \@tempc \@empty \let \@tempc \@tempb \let \@tempb \@tempa \fi \ifx \@tempb \@empty \def\@tempb {arXiv}\fi \@ifundefined {mn@eprint@\@tempb}{\@tempb:\@tempc}{\expandafter \expandafter \csname mn@eprint@\@tempb\endcsname \expandafter{\@tempc}}}

\bibitem[\protect\citeauthoryear{{Abdurro'uf} et~al.,}{{Abdurro'uf} et~al.}{2022}]{sdssdr17}
{Abdurro'uf} et~al., 2022, \mn@doi [\apjs] {10.3847/1538-4365/ac4414}, \href {https://ui.adsabs.harvard.edu/abs/2022ApJS..259...35A} {259, 35}

\bibitem[\protect\citeauthoryear{{Almaini} et~al.,}{{Almaini} et~al.}{2017}]{almaini2017}
{Almaini} O.,  et~al., 2017, \mn@doi [\mnras] {10.1093/mnras/stx1957}, \href {https://ui.adsabs.harvard.edu/abs/2017MNRAS.472.1401A} {472, 1401}

\bibitem[\protect\citeauthoryear{{Alonso}, {Mesa}, {Padilla}  \& {Lambas}}{{Alonso} et~al.}{2012}]{alonso2012}
{Alonso} S.,  {Mesa} V.,  {Padilla} N.,   {Lambas} D.~G.,  2012, \mn@doi [\aap] {10.1051/0004-6361/201117901}, \href {https://ui.adsabs.harvard.edu/abs/2012A&A...539A..46A} {539, A46}

\bibitem[\protect\citeauthoryear{{Andrews} \& {Martini}}{{Andrews} \& {Martini}}{2013}]{andrews2013}
{Andrews} B.~H.,  {Martini} P.,  2013, \mn@doi [\apj] {10.1088/0004-637X/765/2/140}, \href {https://ui.adsabs.harvard.edu/abs/2013ApJ...765..140A} {765, 140}

\bibitem[\protect\citeauthoryear{{Asplund}, {Grevesse}, {Sauval}, {Allende Prieto}  \& {Kiselman}}{{Asplund} et~al.}{2004}]{asplund2004}
{Asplund} M.,  {Grevesse} N.,  {Sauval} A.~J.,  {Allende Prieto} C.,   {Kiselman} D.,  2004, \mn@doi [\aap] {10.1051/0004-6361:20034328}, \href {https://ui.adsabs.harvard.edu/abs/2004A&A...417..751A} {417, 751}

\bibitem[\protect\citeauthoryear{{Asplund}, {Grevesse}, {Sauval}  \& {Scott}}{{Asplund} et~al.}{2009}]{asplund2009}
{Asplund} M.,  {Grevesse} N.,  {Sauval} A.~J.,   {Scott} P.,  2009, \mn@doi [\araa] {10.1146/annurev.astro.46.060407.145222}, \href {https://ui.adsabs.harvard.edu/abs/2009ARA&A..47..481A} {47, 481}

\bibitem[\protect\citeauthoryear{{Baker} \& {Maiolino}}{{Baker} \& {Maiolino}}{2023}]{baker2023a}
{Baker} W.~M.,  {Maiolino} R.,  2023, \mn@doi [\mnras] {10.1093/mnras/stad802}, \href {https://ui.adsabs.harvard.edu/abs/2023MNRAS.521.4173B} {521, 4173}

\bibitem[\protect\citeauthoryear{{Baker} et~al.,}{{Baker} et~al.}{2023}]{baker2023}
{Baker} W.~M.,  et~al., 2023, \mn@doi [\mnras] {10.1093/mnras/stac3594}, \href {https://ui.adsabs.harvard.edu/abs/2023MNRAS.519.1149B} {519, 1149}

\bibitem[\protect\citeauthoryear{{Baldwin}, {Phillips}  \& {Terlevich}}{{Baldwin} et~al.}{1981}]{bpt}
{Baldwin} J.~A.,  {Phillips} M.~M.,   {Terlevich} R.,  1981, \mn@doi [\pasp] {10.1086/130766}, \href {http://adsabs.harvard.edu/abs/1981PASP...93....5B} {93, 5}

\bibitem[\protect\citeauthoryear{{Barone} et~al.,}{{Barone} et~al.}{2018}]{barone2018}
{Barone} T.~M.,  et~al., 2018, \mn@doi [\apj] {10.3847/1538-4357/aaaf6e}, \href {https://ui.adsabs.harvard.edu/abs/2018ApJ...856...64B} {856, 64}

\bibitem[\protect\citeauthoryear{{Barone}, {D'Eugenio}, {Colless}  \& {Scott}}{{Barone} et~al.}{2020}]{barone2020}
{Barone} T.~M.,  {D'Eugenio} F.,  {Colless} M.,   {Scott} N.,  2020, \mn@doi [\apj] {10.3847/1538-4357/ab9951}, \href {https://ui.adsabs.harvard.edu/abs/2020ApJ...898...62B} {898, 62}

\bibitem[\protect\citeauthoryear{{Barrera-Ballesteros}, {S{\'a}nchez}, {Heckman}, {Blanc}  \& {The MaNGA Team}}{{Barrera-Ballesteros} et~al.}{2017}]{bb2017}
{Barrera-Ballesteros} J.~K.,  {S{\'a}nchez} S.~F.,  {Heckman} T.,  {Blanc} G.~A.,   {The MaNGA Team} 2017, \mn@doi [\apj] {10.3847/1538-4357/aa7aa9}, \href {https://ui.adsabs.harvard.edu/abs/2017ApJ...844...80B} {844, 80}

\bibitem[\protect\citeauthoryear{{Barrera-Ballesteros} et~al.,}{{Barrera-Ballesteros} et~al.}{2018}]{bb2018}
{Barrera-Ballesteros} J.~K.,  et~al., 2018, \mn@doi [\apj] {10.3847/1538-4357/aa9b31}, \href {https://ui.adsabs.harvard.edu/abs/2018ApJ...852...74B} {852, 74}

\bibitem[\protect\citeauthoryear{{Belfiore} et~al.,}{{Belfiore} et~al.}{2019}]{belfiore2019}
{Belfiore} F.,  et~al., 2019, \mn@doi [\aj] {10.3847/1538-3881/ab3e4e}, \href {https://ui.adsabs.harvard.edu/abs/2019AJ....158..160B} {158, 160}

\bibitem[\protect\citeauthoryear{{Blanton}, {Kazin}, {Muna}, {Weaver}  \& {Price-Whelan}}{{Blanton} et~al.}{2011}]{blanton2011}
{Blanton} M.~R.,  {Kazin} E.,  {Muna} D.,  {Weaver} B.~A.,   {Price-Whelan} A.,  2011, \mn@doi [\aj] {10.1088/0004-6256/142/1/31}, \href {http://adsabs.harvard.edu/abs/2011AJ....142...31B} {142, 31}

\bibitem[\protect\citeauthoryear{{Blanton} et~al.,}{{Blanton} et~al.}{2017}]{blanton2017}
{Blanton} M.~R.,  et~al., 2017, \mn@doi [\aj] {10.3847/1538-3881/aa7567}, \href {http://adsabs.harvard.edu/abs/2017AJ....154...28B} {154, 28}

\bibitem[\protect\citeauthoryear{{Boardman}, {Zasowski}, {Newman}, {Sanchez}, {Schaefer}, {Lian}, {Bizyaev}  \& {Drory}}{{Boardman} et~al.}{2021}]{boardman2021}
{Boardman} N.~F.,  {Zasowski} G.,  {Newman} J.~A.,  {Sanchez} S.~F.,  {Schaefer} A.,  {Lian} J.,  {Bizyaev} D.,   {Drory} N.,  2021, \mn@doi [\mnras] {10.1093/mnras/staa3785}, \href {https://ui.adsabs.harvard.edu/abs/2021MNRAS.501..948B} {501, 948}

\bibitem[\protect\citeauthoryear{{Boardman} et~al.,}{{Boardman} et~al.}{2022}]{boardman2022}
{Boardman} N.,  et~al., 2022, \mn@doi [\mnras] {10.1093/mnras/stac1475}, \href {https://ui.adsabs.harvard.edu/abs/2022MNRAS.514.2298B} {514, 2298}

\bibitem[\protect\citeauthoryear{{Boardman}, {Wild}, {Heckman}, {Sanchez}, {Riffel}, {Riffel}  \& {Zasowski}}{{Boardman} et~al.}{2023}]{boardman2023}
{Boardman} N.,  {Wild} V.,  {Heckman} T.,  {Sanchez} S.~F.,  {Riffel} R.,  {Riffel} R.~A.,   {Zasowski} G.,  2023, \mn@doi [\mnras] {10.1093/mnras/stad277}, \href {https://ui.adsabs.harvard.edu/abs/2023MNRAS.520.4301B} {520, 4301}

\bibitem[\protect\citeauthoryear{{Boardman}, {Wild}, {Rowlands}, {Vale Asari}  \& {Luo}}{{Boardman} et~al.}{2024a}]{boardman2024}
{Boardman} N.,  {Wild} V.,  {Rowlands} K.,  {Vale Asari} N.,   {Luo} Y.,  2024a, \mn@doi [\mnras] {10.1093/mnras/stad3932}, \href {https://ui.adsabs.harvard.edu/abs/2024MNRAS.52710788B} {527, 10788}

\bibitem[\protect\citeauthoryear{{Boardman}, {Wild}  \& {Asari}}{{Boardman} et~al.}{2024b}]{boardman2024a}
{Boardman} N.,  {Wild} V.,   {Asari} N.~V.,  2024b, \mn@doi [\mnras] {10.1093/mnrasl/slae071}, \href {https://ui.adsabs.harvard.edu/abs/2024MNRAS.534L...1B} {534, L1}

\bibitem[\protect\citeauthoryear{{Bothwell}, {Maiolino}, {Kennicutt}, {Cresci}, {Mannucci}, {Marconi}  \& {Cicone}}{{Bothwell} et~al.}{2013}]{bothwell2013}
{Bothwell} M.~S.,  {Maiolino} R.,  {Kennicutt} R.,  {Cresci} G.,  {Mannucci} F.,  {Marconi} A.,   {Cicone} C.,  2013, \mn@doi [\mnras] {10.1093/mnras/stt817}, \href {https://ui.adsabs.harvard.edu/abs/2013MNRAS.433.1425B} {433, 1425}

\bibitem[\protect\citeauthoryear{{Bothwell}, {Maiolino}, {Cicone}, {Peng}  \& {Wagg}}{{Bothwell} et~al.}{2016}]{bothwell2016}
{Bothwell} M.~S.,  {Maiolino} R.,  {Cicone} C.,  {Peng} Y.,   {Wagg} J.,  2016, \mn@doi [\aap] {10.1051/0004-6361/201527918}, \href {https://ui.adsabs.harvard.edu/abs/2016A&A...595A..48B} {595, A48}

\bibitem[\protect\citeauthoryear{{Brown}, {Cortese}, {Catinella}  \& {Kilborn}}{{Brown} et~al.}{2018}]{brown2018}
{Brown} T.,  {Cortese} L.,  {Catinella} B.,   {Kilborn} V.,  2018, \mn@doi [\mnras] {10.1093/mnras/stx2452}, \href {https://ui.adsabs.harvard.edu/abs/2018MNRAS.473.1868B} {473, 1868}

\bibitem[\protect\citeauthoryear{{Bundy} et~al.,}{{Bundy} et~al.}{2015}]{bundy2015}
{Bundy} K.,  et~al., 2015, \mn@doi [\apj] {10.1088/0004-637X/798/1/7}, \href {http://adsabs.harvard.edu/abs/2015ApJ...798....7B} {798, 7}

\bibitem[\protect\citeauthoryear{{Calura}, {Pipino}, {Chiappini}, {Matteucci}  \& {Maiolino}}{{Calura} et~al.}{2009}]{calura2009}
{Calura} F.,  {Pipino} A.,  {Chiappini} C.,  {Matteucci} F.,   {Maiolino} R.,  2009, \mn@doi [\aap] {10.1051/0004-6361/200911756}, \href {https://ui.adsabs.harvard.edu/abs/2009A&A...504..373C} {504, 373}

\bibitem[\protect\citeauthoryear{{Cappellari} et~al.,}{{Cappellari} et~al.}{2013}]{cappellari2013a}
{Cappellari} M.,  et~al., 2013, \mn@doi [\mnras] {10.1093/mnras/stt644}, \href {http://adsabs.harvard.edu/abs/2013MNRAS.432.1862C} {432, 1862}

\bibitem[\protect\citeauthoryear{{Cardelli}, {Clayton}  \& {Mathis}}{{Cardelli} et~al.}{1989}]{cardelli1989}
{Cardelli} J.~A.,  {Clayton} G.~C.,   {Mathis} J.~S.,  1989, \mn@doi [\apj] {10.1086/167900}, \href {https://ui.adsabs.harvard.edu/abs/1989ApJ...345..245C} {345, 245}

\bibitem[\protect\citeauthoryear{{Chabrier}}{{Chabrier}}{2003}]{chabrier2003}
{Chabrier} G.,  2003, \mn@doi [\pasp] {10.1086/376392}, \href {http://adsabs.harvard.edu/abs/2003PASP..115..763C} {115, 763}

\bibitem[\protect\citeauthoryear{{Chen}, {Lin}, {Kong}, {Liang}, {Chen}  \& {Zhang}}{{Chen} et~al.}{2022}]{chen2022}
{Chen} X.,  {Lin} Z.,  {Kong} X.,  {Liang} Z.,  {Chen} G.,   {Zhang} H.-X.,  2022, \mn@doi [\apj] {10.3847/1538-4357/ac75b4}, \href {https://ui.adsabs.harvard.edu/abs/2022ApJ...933..228C} {933, 228}

\bibitem[\protect\citeauthoryear{{Cherinka} et~al.,}{{Cherinka} et~al.}{2019}]{cherinka2019}
{Cherinka} B.,  et~al., 2019, \mn@doi [\aj] {10.3847/1538-3881/ab2634}, \href {https://ui.adsabs.harvard.edu/abs/2019AJ....158...74C} {158, 74}

\bibitem[\protect\citeauthoryear{{Chiappini}, {Matteucci}  \& {Ballero}}{{Chiappini} et~al.}{2005}]{chiappini2005}
{Chiappini} C.,  {Matteucci} F.,   {Ballero} S.~K.,  2005, \mn@doi [\aap] {10.1051/0004-6361:20042292}, \href {https://ui.adsabs.harvard.edu/abs/2005A&A...437..429C} {437, 429}

\bibitem[\protect\citeauthoryear{{Cirasuolo} et~al.,}{{Cirasuolo} et~al.}{2020}]{cirasuolo2020}
{Cirasuolo} M.,  et~al., 2020, \mn@doi [The Messenger] {10.18727/0722-6691/5195}, \href {https://ui.adsabs.harvard.edu/abs/2020Msngr.180...10C} {180, 10}

\bibitem[\protect\citeauthoryear{Cleveland \& Devlin}{Cleveland \& Devlin}{1988}]{cleveland1988}
Cleveland W.~S.,  Devlin S.~J.,  1988, \mn@doi [Journal of the American Statistical Association] {10.1080/01621459.1988.10478639}, 83, 596

\bibitem[\protect\citeauthoryear{{Conroy}}{{Conroy}}{2013}]{conroy2013}
{Conroy} C.,  2013, \mn@doi [\araa] {10.1146/annurev-astro-082812-141017}, \href {https://ui.adsabs.harvard.edu/abs/2013ARA&A..51..393C} {51, 393}

\bibitem[\protect\citeauthoryear{{Cresci}, {Mannucci}, {Sommariva}, {Maiolino}, {Marconi}  \& {Brusa}}{{Cresci} et~al.}{2012}]{cresci2012}
{Cresci} G.,  {Mannucci} F.,  {Sommariva} V.,  {Maiolino} R.,  {Marconi} A.,   {Brusa} M.,  2012, \mn@doi [\mnras] {10.1111/j.1365-2966.2011.20299.x}, \href {https://ui.adsabs.harvard.edu/abs/2012MNRAS.421..262C} {421, 262}

\bibitem[\protect\citeauthoryear{{Cresci}, {Mannucci}  \& {Curti}}{{Cresci} et~al.}{2019}]{cresci2019}
{Cresci} G.,  {Mannucci} F.,   {Curti} M.,  2019, \mn@doi [\aap] {10.1051/0004-6361/201834637}, \href {https://ui.adsabs.harvard.edu/abs/2019A&A...627A..42C} {627, A42}

\bibitem[\protect\citeauthoryear{{Curti}, {Mannucci}, {Cresci}  \& {Maiolino}}{{Curti} et~al.}{2020}]{curti2020}
{Curti} M.,  {Mannucci} F.,  {Cresci} G.,   {Maiolino} R.,  2020, \mn@doi [\mnras] {10.1093/mnras/stz2910}, \href {https://ui.adsabs.harvard.edu/abs/2020MNRAS.491..944C} {491, 944}

\bibitem[\protect\citeauthoryear{{Curti} et~al.,}{{Curti} et~al.}{2023}]{curti2023}
{Curti} M.,  et~al., 2023, \mn@doi [\mnras] {10.1093/mnras/stac2737}, \href {https://ui.adsabs.harvard.edu/abs/2023MNRAS.518..425C} {518, 425}

\bibitem[\protect\citeauthoryear{{Curti} et~al.,}{{Curti} et~al.}{2024}]{curti2024}
{Curti} M.,  et~al., 2024, \mn@doi [\aap] {10.1051/0004-6361/202346698}, \href {https://ui.adsabs.harvard.edu/abs/2024A&A...684A..75C} {684, A75}

\bibitem[\protect\citeauthoryear{{D'Eugenio}, {Colless}, {Groves}, {Bian}  \& {Barone}}{{D'Eugenio} et~al.}{2018}]{deugenio2018}
{D'Eugenio} F.,  {Colless} M.,  {Groves} B.,  {Bian} F.,   {Barone} T.~M.,  2018, \mn@doi [\mnras] {10.1093/mnras/sty1424}, \href {https://ui.adsabs.harvard.edu/abs/2018MNRAS.479.1807D} {479, 1807}

\bibitem[\protect\citeauthoryear{{Dav{\'e}}, {Finlator}  \& {Oppenheimer}}{{Dav{\'e}} et~al.}{2012}]{dave2012}
{Dav{\'e}} R.,  {Finlator} K.,   {Oppenheimer} B.~D.,  2012, \mn@doi [\mnras] {10.1111/j.1365-2966.2011.20148.x}, \href {https://ui.adsabs.harvard.edu/abs/2012MNRAS.421...98D} {421, 98}

\bibitem[\protect\citeauthoryear{{Dav{\'e}}, {Thompson}  \& {Hopkins}}{{Dav{\'e}} et~al.}{2016}]{dave2016}
{Dav{\'e}} R.,  {Thompson} R.,   {Hopkins} P.~F.,  2016, \mn@doi [\mnras] {10.1093/mnras/stw1862}, \href {https://ui.adsabs.harvard.edu/abs/2016MNRAS.462.3265D} {462, 3265}

\bibitem[\protect\citeauthoryear{{Dav{\'e}}, {Rafieferantsoa}, {Thompson}  \& {Hopkins}}{{Dav{\'e}} et~al.}{2017}]{dave2017}
{Dav{\'e}} R.,  {Rafieferantsoa} M.~H.,  {Thompson} R.~J.,   {Hopkins} P.~F.,  2017, \mn@doi [\mnras] {10.1093/mnras/stx108}, \href {https://ui.adsabs.harvard.edu/abs/2017MNRAS.467..115D} {467, 115}

\bibitem[\protect\citeauthoryear{{De Rossi}, {Bower}, {Font}, {Schaye}  \& {Theuns}}{{De Rossi} et~al.}{2017}]{derossi2017}
{De Rossi} M.~E.,  {Bower} R.~G.,  {Font} A.~S.,  {Schaye} J.,   {Theuns} T.,  2017, \mn@doi [\mnras] {10.1093/mnras/stx2158}, \href {https://ui.adsabs.harvard.edu/abs/2017MNRAS.472.3354D} {472, 3354}

\bibitem[\protect\citeauthoryear{{De Rossi}, {Bower}, {Font}  \& {Schaye}}{{De Rossi} et~al.}{2018}]{derossi2018}
{De Rossi} M.~E.,  {Bower} R.~G.,  {Font} A.~S.,   {Schaye} T.,  2018, \mn@doi [Boletin de la Asociacion Argentina de Astronomia La Plata Argentina] {10.48550/arXiv.1805.06119}, \href {https://ui.adsabs.harvard.edu/abs/2018BAAA...60..121R} {60, 121}

\bibitem[\protect\citeauthoryear{{Dopita}, {Kewley}, {Sutherland}  \& {Nicholls}}{{Dopita} et~al.}{2016}]{Dopita_2016_EmLineDiagnostic}
{Dopita} M.~A.,  {Kewley} L.~J.,  {Sutherland} R.~S.,   {Nicholls} D.~C.,  2016, \mn@doi [\apss] {10.1007/s10509-016-2657-8}, \href {http://adsabs.harvard.edu/abs/2016Ap%26SS.361...61D} {361, 61}

\bibitem[\protect\citeauthoryear{{Drory} et~al.,}{{Drory} et~al.}{2015}]{drory2015}
{Drory} N.,  et~al., 2015, \mn@doi [\aj] {10.1088/0004-6256/149/2/77}, \href {http://adsabs.harvard.edu/abs/2015AJ....149...77D} {149, 77}

\bibitem[\protect\citeauthoryear{{Duarte Puertas}, {Vilchez}, {Iglesias-P{\'a}ramo}, {Moll{\'a}}, {P{\'e}rez-Montero}, {Kehrig}, {Pilyugin}  \& {Zinchenko}}{{Duarte Puertas} et~al.}{2022}]{dp2022}
{Duarte Puertas} S.,  {Vilchez} J.~M.,  {Iglesias-P{\'a}ramo} J.,  {Moll{\'a}} M.,  {P{\'e}rez-Montero} E.,  {Kehrig} C.,  {Pilyugin} L.~S.,   {Zinchenko} I.~A.,  2022, \mn@doi [\aap] {10.1051/0004-6361/202141571}, \href {https://ui.adsabs.harvard.edu/abs/2022A&A...666A.186D} {666, A186}

\bibitem[\protect\citeauthoryear{{Edmunds} \& {Pagel}}{{Edmunds} \& {Pagel}}{1978}]{edmunds1978}
{Edmunds} M.~G.,  {Pagel} B.~E.~J.,  1978, \mn@doi [\mnras] {10.1093/mnras/185.1.77P}, \href {https://ui.adsabs.harvard.edu/abs/1978MNRAS.185P..77E} {185, 77P}

\bibitem[\protect\citeauthoryear{{Ellison}, {Patton}, {Simard}  \& {McConnachie}}{{Ellison} et~al.}{2008a}]{ellison2008a}
{Ellison} S.~L.,  {Patton} D.~R.,  {Simard} L.,   {McConnachie} A.~W.,  2008a, \mn@doi [\aj] {10.1088/0004-6256/135/5/1877}, \href {https://ui.adsabs.harvard.edu/abs/2008AJ....135.1877E} {135, 1877}

\bibitem[\protect\citeauthoryear{{Ellison}, {Patton}, {Simard}  \& {McConnachie}}{{Ellison} et~al.}{2008b}]{ellison2008}
{Ellison} S.~L.,  {Patton} D.~R.,  {Simard} L.,   {McConnachie} A.~W.,  2008b, \mn@doi [\apjl] {10.1086/527296}, \href {https://ui.adsabs.harvard.edu/abs/2008ApJ...672L.107E} {672, L107}

\bibitem[\protect\citeauthoryear{{Faisst} et~al.,}{{Faisst} et~al.}{2016}]{faisst2016}
{Faisst} A.~L.,  et~al., 2016, \mn@doi [\apj] {10.3847/0004-637X/822/1/29}, \href {https://ui.adsabs.harvard.edu/abs/2016ApJ...822...29F} {822, 29}

\bibitem[\protect\citeauthoryear{{Fernandez}, {Alonso}, {Mesa}  \& {Duplancic}}{{Fernandez} et~al.}{2024}]{fernandez2024}
{Fernandez} J.,  {Alonso} S.,  {Mesa} V.,   {Duplancic} F.,  2024, \mn@doi [\aap] {10.1051/0004-6361/202245215}, \href {https://ui.adsabs.harvard.edu/abs/2024A&A...683A..32F} {683, A32}

\bibitem[\protect\citeauthoryear{{Fitzpatrick}, {Massa}, {Gordon}, {Bohlin}  \& {Clayton}}{{Fitzpatrick} et~al.}{2019}]{fitzpatrick2019}
{Fitzpatrick} E.~L.,  {Massa} D.,  {Gordon} K.~D.,  {Bohlin} R.,   {Clayton} G.~C.,  2019, \mn@doi [\apj] {10.3847/1538-4357/ab4c3a}, \href {https://ui.adsabs.harvard.edu/abs/2019ApJ...886..108F} {886, 108}

\bibitem[\protect\citeauthoryear{{Florido}, {Zurita}  \& {P{\'e}rez-Montero}}{{Florido} et~al.}{2022}]{florido2022}
{Florido} E.,  {Zurita} A.,   {P{\'e}rez-Montero} E.,  2022, \mn@doi [\mnras] {10.1093/mnras/stac919}, \href {https://ui.adsabs.harvard.edu/abs/2022MNRAS.513.2006F} {513, 2006}

\bibitem[\protect\citeauthoryear{{Forbes}, {Krumholz}, {Burkert}  \& {Dekel}}{{Forbes} et~al.}{2014}]{forbes2014}
{Forbes} J.~C.,  {Krumholz} M.~R.,  {Burkert} A.,   {Dekel} A.,  2014, \mn@doi [\mnras] {10.1093/mnras/stu1142}, \href {https://ui.adsabs.harvard.edu/abs/2014MNRAS.443..168F} {443, 168}

\bibitem[\protect\citeauthoryear{{Fraser-McKelvie}, {Brown}, {Pimbblet}, {Dolley}  \& {Bonne}}{{Fraser-McKelvie} et~al.}{2018}]{frazermckelvie2018}
{Fraser-McKelvie} A.,  {Brown} M. J.~I.,  {Pimbblet} K.,  {Dolley} T.,   {Bonne} N.~J.,  2018, \mn@doi [\mnras] {10.1093/mnras/stx2823}, \href {https://ui.adsabs.harvard.edu/abs/2018MNRAS.474.1909F} {474, 1909}

\bibitem[\protect\citeauthoryear{{Fraser-McKelvie}, {Merrifield}  \& {Arag{\'o}n-Salamanca}}{{Fraser-McKelvie} et~al.}{2019}]{frasermckelvie2019b}
{Fraser-McKelvie} A.,  {Merrifield} M.,   {Arag{\'o}n-Salamanca} A.,  2019, \mn@doi [\mnras] {10.1093/mnras/stz2493}, \href {https://ui.adsabs.harvard.edu/abs/2019MNRAS.489.5030F} {489, 5030}

\bibitem[\protect\citeauthoryear{{Fraser-McKelvie} et~al.,}{{Fraser-McKelvie} et~al.}{2022}]{frasermckelvie2022}
{Fraser-McKelvie} A.,  et~al., 2022, \mn@doi [\mnras] {10.1093/mnras/stab3430}, \href {https://ui.adsabs.harvard.edu/abs/2022MNRAS.510..320F} {510, 320}

\bibitem[\protect\citeauthoryear{{Gallazzi}, {Charlot}, {Brinchmann}, {White}  \& {Tremonti}}{{Gallazzi} et~al.}{2005}]{gallazzi2005}
{Gallazzi} A.,  {Charlot} S.,  {Brinchmann} J.,  {White} S. D.~M.,   {Tremonti} C.~A.,  2005, \mn@doi [\mnras] {10.1111/j.1365-2966.2005.09321.x}, \href {https://ui.adsabs.harvard.edu/abs/2005MNRAS.362...41G} {362, 41}

\bibitem[\protect\citeauthoryear{{Gallazzi}, {Charlot}, {Brinchmann}  \& {White}}{{Gallazzi} et~al.}{2006}]{gallazzi2006}
{Gallazzi} A.,  {Charlot} S.,  {Brinchmann} J.,   {White} S. D.~M.,  2006, \mn@doi [\mnras] {10.1111/j.1365-2966.2006.10548.x}, \href {https://ui.adsabs.harvard.edu/abs/2006MNRAS.370.1106G} {370, 1106}

\bibitem[\protect\citeauthoryear{{Garc{\'\i}a-Benito} et~al.,}{{Garc{\'\i}a-Benito} et~al.}{2017}]{gb2017}
{Garc{\'\i}a-Benito} R.,  et~al., 2017, \mn@doi [\aap] {10.1051/0004-6361/201731357}, \href {https://ui.adsabs.harvard.edu/abs/2017A&A...608A..27G} {608, A27}

\bibitem[\protect\citeauthoryear{{Garcia} et~al.,}{{Garcia} et~al.}{2024}]{garcia2024a}
{Garcia} A.~M.,  et~al., 2024, \mn@doi [\mnras] {10.1093/mnras/stae737}, \href {https://ui.adsabs.harvard.edu/abs/2024MNRAS.529.3342G} {529, 3342}

\bibitem[\protect\citeauthoryear{{Greene}, {Bezanson}, {Ouchi}, {Silverman}  \& {the PFS Galaxy Evolution Working Group}}{{Greene} et~al.}{2022}]{greene2022}
{Greene} J.,  {Bezanson} R.,  {Ouchi} M.,  {Silverman} J.,   {the PFS Galaxy Evolution Working Group} 2022, \mn@doi [arXiv e-prints] {10.48550/arXiv.2206.14908}, \href {https://ui.adsabs.harvard.edu/abs/2022arXiv220614908G} {p. arXiv:2206.14908}

\bibitem[\protect\citeauthoryear{{Greener} et~al.,}{{Greener} et~al.}{2022}]{greener2022}
{Greener} M.~J.,  et~al., 2022, \mn@doi [\mnras] {10.1093/mnras/stac2355}, \href {https://ui.adsabs.harvard.edu/abs/2022MNRAS.516.1275G} {516, 1275}

\bibitem[\protect\citeauthoryear{{Gunn} et~al.,}{{Gunn} et~al.}{2006}]{gunn2006}
{Gunn} J.~E.,  et~al., 2006, \mn@doi [\aj] {10.1086/500975}, \href {http://adsabs.harvard.edu/abs/2006AJ....131.2332G} {131, 2332}

\bibitem[\protect\citeauthoryear{{Hayden-Pawson} et~al.,}{{Hayden-Pawson} et~al.}{2022}]{hp2022}
{Hayden-Pawson} C.,  et~al., 2022, \mn@doi [\mnras] {10.1093/mnras/stac584}, \href {https://ui.adsabs.harvard.edu/abs/2022MNRAS.512.2867H} {512, 2867}

\bibitem[\protect\citeauthoryear{{Heintz} et~al.,}{{Heintz} et~al.}{2023}]{heintz2023}
{Heintz} K.~E.,  et~al., 2023, \mn@doi [Nature Astronomy] {10.1038/s41550-023-02078-7}, \href {https://ui.adsabs.harvard.edu/abs/2023NatAs...7.1517H} {7, 1517}

\bibitem[\protect\citeauthoryear{{Jara-Ferreira}, {Tissera}, {Sillero}, {Rosas-Guevara}, {Pedrosa}, {De Rossi}, {Theuns}  \& {Bignone}}{{Jara-Ferreira} et~al.}{2024}]{jf2024}
{Jara-Ferreira} F.,  {Tissera} P.~B.,  {Sillero} E.,  {Rosas-Guevara} Y.,  {Pedrosa} S.~E.,  {De Rossi} M.~E.,  {Theuns} T.,   {Bignone} L.,  2024, \mn@doi [\mnras] {10.1093/mnras/stae708}, \href {https://ui.adsabs.harvard.edu/abs/2024MNRAS.530.1369J} {530, 1369}

\bibitem[\protect\citeauthoryear{{Johnson} et~al.,}{{Johnson} et~al.}{2021}]{johnson2021}
{Johnson} J.~W.,  et~al., 2021, \mn@doi [\mnras] {10.1093/mnras/stab2718}, \href {https://ui.adsabs.harvard.edu/abs/2021MNRAS.508.4484J} {508, 4484}

\bibitem[\protect\citeauthoryear{{Johnson}, {Weinberg}, {Vincenzo}, {Bird}  \& {Griffith}}{{Johnson} et~al.}{2023}]{johnson2023}
{Johnson} J.~W.,  {Weinberg} D.~H.,  {Vincenzo} F.,  {Bird} J.~C.,   {Griffith} E.~J.,  2023, \mn@doi [\mnras] {10.1093/mnras/stad057}, \href {https://ui.adsabs.harvard.edu/abs/2023MNRAS.520..782J} {520, 782}

\bibitem[\protect\citeauthoryear{{Jones}, {Sanders}, {Roberts-Borsani}, {Ellis}, {Laporte}, {Treu}  \& {Harikane}}{{Jones} et~al.}{2020}]{jones2020}
{Jones} T.,  {Sanders} R.,  {Roberts-Borsani} G.,  {Ellis} R.~S.,  {Laporte} N.,  {Treu} T.,   {Harikane} Y.,  2020, \mn@doi [\apj] {10.3847/1538-4357/abb943}, \href {https://ui.adsabs.harvard.edu/abs/2020ApJ...903..150J} {903, 150}

\bibitem[\protect\citeauthoryear{{Kashino}, {Renzini}, {Silverman}  \& {Daddi}}{{Kashino} et~al.}{2016}]{kashino2016}
{Kashino} D.,  {Renzini} A.,  {Silverman} J.~D.,   {Daddi} E.,  2016, \mn@doi [\apjl] {10.3847/2041-8205/823/2/L24}, \href {https://ui.adsabs.harvard.edu/abs/2016ApJ...823L..24K} {823, L24}

\bibitem[\protect\citeauthoryear{{Kauffmann} et~al.,}{{Kauffmann} et~al.}{2003}]{kauffmann2003}
{Kauffmann} G.,  et~al., 2003, \mn@doi [\mnras] {10.1111/j.1365-2966.2003.07154.x}, \href {http://adsabs.harvard.edu/abs/2003MNRAS.346.1055K} {346, 1055}

\bibitem[\protect\citeauthoryear{{Kewley}, {Heisler}, {Dopita}  \& {Lumsden}}{{Kewley} et~al.}{2001}]{kewley2001}
{Kewley} L.~J.,  {Heisler} C.~A.,  {Dopita} M.~A.,   {Lumsden} S.,  2001, \mn@doi [\apjs] {10.1086/318944}, \href {http://adsabs.harvard.edu/abs/2001ApJS..132...37K} {132, 37}

\bibitem[\protect\citeauthoryear{{K{\"o}ppen} \& {Hensler}}{{K{\"o}ppen} \& {Hensler}}{2005}]{koppen2005}
{K{\"o}ppen} J.,  {Hensler} G.,  2005, \mn@doi [\aap] {10.1051/0004-6361:20042266}, \href {https://ui.adsabs.harvard.edu/abs/2005A&A...434..531K} {434, 531}

\bibitem[\protect\citeauthoryear{{Lacerda} et~al.,}{{Lacerda} et~al.}{2018}]{lacerda2018}
{Lacerda} E.~A.~D.,  et~al., 2018, \mn@doi [\mnras] {10.1093/mnras/stx3022}, \href {https://ui.adsabs.harvard.edu/abs/2018MNRAS.474.3727L} {474, 3727}

\bibitem[\protect\citeauthoryear{{Lagos} et~al.,}{{Lagos} et~al.}{2016}]{lagos2016}
{Lagos} C. d.~P.,  et~al., 2016, \mn@doi [\mnras] {10.1093/mnras/stw717}, \href {https://ui.adsabs.harvard.edu/abs/2016MNRAS.459.2632L} {459, 2632}

\bibitem[\protect\citeauthoryear{{Lara-L{\'o}pez} et~al.,}{{Lara-L{\'o}pez} et~al.}{2010}]{ll2010}
{Lara-L{\'o}pez} M.~A.,  et~al., 2010, \mn@doi [\aap] {10.1051/0004-6361/201014803}, \href {https://ui.adsabs.harvard.edu/abs/2010A&A...521L..53L} {521, L53}

\bibitem[\protect\citeauthoryear{{Law} et~al.,}{{Law} et~al.}{2015}]{law2015}
{Law} D.~R.,  et~al., 2015, \mn@doi [\aj] {10.1088/0004-6256/150/1/19}, \href {http://adsabs.harvard.edu/abs/2015AJ....150...19L} {150, 19}

\bibitem[\protect\citeauthoryear{{Law} et~al.,}{{Law} et~al.}{2016}]{law2016}
{Law} D.~R.,  et~al., 2016, \mn@doi [\aj] {10.3847/0004-6256/152/4/83}, \href {https://ui.adsabs.harvard.edu/\#abs/2016AJ....152...83L} {152, 83}

\bibitem[\protect\citeauthoryear{{Law} et~al.,}{{Law} et~al.}{2021}]{law2021}
{Law} D.~R.,  et~al., 2021, \mn@doi [\aj] {10.3847/1538-3881/abcaa2}, \href {https://ui.adsabs.harvard.edu/abs/2021AJ....161...52L} {161, 52}

\bibitem[\protect\citeauthoryear{{Lequeux}, {Peimbert}, {Rayo}, {Serrano}  \& {Torres-Peimbert}}{{Lequeux} et~al.}{1979}]{lequeux1979}
{Lequeux} J.,  {Peimbert} M.,  {Rayo} J.~F.,  {Serrano} A.,   {Torres-Peimbert} S.,  1979, \aap, \href {https://ui.adsabs.harvard.edu/abs/1979A&A....80..155L} {500, 145}

\bibitem[\protect\citeauthoryear{{Leung}, {Wild}, {Papathomas}, {Carnall}, {Zheng}, {Boardman}, {Wang}  \& {Johansson}}{{Leung} et~al.}{2024}]{leung2024}
{Leung} H.-H.,  {Wild} V.,  {Papathomas} M.,  {Carnall} A.,  {Zheng} Y.,  {Boardman} N.,  {Wang} C.,   {Johansson} P.~H.,  2024, \mn@doi [\mnras] {10.1093/mnras/stae225}, \href {https://ui.adsabs.harvard.edu/abs/2024MNRAS.528.4029L} {528, 4029}

\bibitem[\protect\citeauthoryear{{Lewis} et~al.,}{{Lewis} et~al.}{2024}]{lewis2024}
{Lewis} Z.~J.,  et~al., 2024, \mn@doi [\apj] {10.3847/1538-4357/ad250c}, \href {https://ui.adsabs.harvard.edu/abs/2024ApJ...964...59L} {964, 59}

\bibitem[\protect\citeauthoryear{{Li} et~al.,}{{Li} et~al.}{2018}]{li2018}
{Li} H.,  et~al., 2018, \mn@doi [\mnras] {10.1093/mnras/sty334}, \href {http://adsabs.harvard.edu/abs/2018MNRAS.476.1765L} {476, 1765}

\bibitem[\protect\citeauthoryear{{Lian}, {Li}, {Yan}  \& {Kong}}{{Lian} et~al.}{2015}]{lian2015}
{Lian} J.~H.,  {Li} J.~R.,  {Yan} W.,   {Kong} X.,  2015, \mn@doi [\mnras] {10.1093/mnras/stu2184}, \href {https://ui.adsabs.harvard.edu/abs/2015MNRAS.446.1449L} {446, 1449}

\bibitem[\protect\citeauthoryear{{Lian}, {Thomas}, {Maraston}, {Goddard}, {Comparat}, {Gonzalez-Perez}  \& {Ventura}}{{Lian} et~al.}{2018}]{lian2018a}
{Lian} J.,  {Thomas} D.,  {Maraston} C.,  {Goddard} D.,  {Comparat} J.,  {Gonzalez-Perez} V.,   {Ventura} P.,  2018, \mn@doi [\mnras] {10.1093/mnras/stx2829}, \href {https://ui.adsabs.harvard.edu/abs/2018MNRAS.474.1143L} {474, 1143}

\bibitem[\protect\citeauthoryear{{Lilly}, {Carollo}, {Pipino}, {Renzini}  \& {Peng}}{{Lilly} et~al.}{2013}]{lilly2013}
{Lilly} S.~J.,  {Carollo} C.~M.,  {Pipino} A.,  {Renzini} A.,   {Peng} Y.,  2013, \mn@doi [\apj] {10.1088/0004-637X/772/2/119}, \href {https://ui.adsabs.harvard.edu/abs/2013ApJ...772..119L} {772, 119}

\bibitem[\protect\citeauthoryear{{Looser}, {D'Eugenio}, {Piotrowska}, {Belfiore}, {Maiolino}, {Cappellari}, {Baker}  \& {Tacchella}}{{Looser} et~al.}{2024}]{looser2024}
{Looser} T.~J.,  {D'Eugenio} F.,  {Piotrowska} J.~M.,  {Belfiore} F.,  {Maiolino} R.,  {Cappellari} M.,  {Baker} W.~M.,   {Tacchella} S.,  2024, \mn@doi [\mnras] {10.1093/mnras/stae1581}, \href {https://ui.adsabs.harvard.edu/abs/2024MNRAS.532.2832L} {532, 2832}

\bibitem[\protect\citeauthoryear{{L{\'o}pez Fern{\'a}ndez} et~al.,}{{L{\'o}pez Fern{\'a}ndez} et~al.}{2016}]{lopezfernandez2016}
{L{\'o}pez Fern{\'a}ndez} R.,  et~al., 2016, \mn@doi [\mnras] {10.1093/mnras/stw260}, \href {https://ui.adsabs.harvard.edu/abs/2016MNRAS.458..184L} {458, 184}

\bibitem[\protect\citeauthoryear{{Ma}, {Du}, {Ho}, {Sheng}  \& {Liao}}{{Ma} et~al.}{2024a}]{ma2024}
{Ma} H.-C.,  {Du} M.,  {Ho} L.~C.,  {Sheng} M.-J.,   {Liao} S.,  2024a, \mn@doi [\aap] {10.1051/0004-6361/202450397}, \href {https://ui.adsabs.harvard.edu/abs/2024A&A...689A.293M} {689, A293}

\bibitem[\protect\citeauthoryear{{Ma} et~al.,}{{Ma} et~al.}{2024b}]{ma2024a}
{Ma} C.,  et~al., 2024b, \mn@doi [\apjl] {10.3847/2041-8213/ad675f}, \href {https://ui.adsabs.harvard.edu/abs/2024ApJ...971L..14M} {971, L14}

\bibitem[\protect\citeauthoryear{{Maiolino} \& {Mannucci}}{{Maiolino} \& {Mannucci}}{2019}]{maiolino2019}
{Maiolino} R.,  {Mannucci} F.,  2019, \mn@doi [\aapr] {10.1007/s00159-018-0112-2}, \href {https://ui.adsabs.harvard.edu/abs/2019A&ARv..27....3M} {27, 3}

\bibitem[\protect\citeauthoryear{{Maiolino} et~al.,}{{Maiolino} et~al.}{2008}]{maiolino2008}
{Maiolino} R.,  et~al., 2008, \mn@doi [\aap] {10.1051/0004-6361:200809678}, \href {https://ui.adsabs.harvard.edu/abs/2008A&A...488..463M} {488, 463}

\bibitem[\protect\citeauthoryear{{Maiolino} et~al.,}{{Maiolino} et~al.}{2020}]{maiolino2020}
{Maiolino} R.,  et~al., 2020, \mn@doi [The Messenger] {10.18727/0722-6691/5197}, \href {https://ui.adsabs.harvard.edu/abs/2020Msngr.180...24M} {180, 24}

\bibitem[\protect\citeauthoryear{{Mannucci}, {Cresci}, {Maiolino}, {Marconi}  \& {Gnerucci}}{{Mannucci} et~al.}{2010}]{mannucci2010}
{Mannucci} F.,  {Cresci} G.,  {Maiolino} R.,  {Marconi} A.,   {Gnerucci} A.,  2010, \mn@doi [\mnras] {10.1111/j.1365-2966.2010.17291.x}, \href {https://ui.adsabs.harvard.edu/abs/2010MNRAS.408.2115M} {408, 2115}

\bibitem[\protect\citeauthoryear{{Maoz}, {Sharon}  \& {Gal-Yam}}{{Maoz} et~al.}{2010}]{maoz2010}
{Maoz} D.,  {Sharon} K.,   {Gal-Yam} A.,  2010, \mn@doi [\apj] {10.1088/0004-637X/722/2/1879}, \href {https://ui.adsabs.harvard.edu/abs/2010ApJ...722.1879M} {722, 1879}

\bibitem[\protect\citeauthoryear{{Marinacci} et~al.,}{{Marinacci} et~al.}{2018}]{tng3}
{Marinacci} F.,  et~al., 2018, \mn@doi [\mnras] {10.1093/mnras/sty2206}, \href {https://ui.adsabs.harvard.edu/abs/2018MNRAS.480.5113M} {480, 5113}

\bibitem[\protect\citeauthoryear{{Marszewski}, {Sun}, {Faucher-Gigu{\`e}re}, {Hayward}  \& {Feldmann}}{{Marszewski} et~al.}{2024}]{marszewski2024}
{Marszewski} A.,  {Sun} G.,  {Faucher-Gigu{\`e}re} C.-A.,  {Hayward} C.~C.,   {Feldmann} R.,  2024, \mn@doi [\apjl] {10.3847/2041-8213/ad4cee}, \href {https://ui.adsabs.harvard.edu/abs/2024ApJ...967L..41M} {967, L41}

\bibitem[\protect\citeauthoryear{{Matteucci}}{{Matteucci}}{1986}]{matteucci1986}
{Matteucci} F.,  1986, \mn@doi [\mnras] {10.1093/mnras/221.4.911}, \href {https://ui.adsabs.harvard.edu/abs/1986MNRAS.221..911M} {221, 911}

\bibitem[\protect\citeauthoryear{{Matthee} \& {Schaye}}{{Matthee} \& {Schaye}}{2018}]{matthee2018}
{Matthee} J.,  {Schaye} J.,  2018, \mn@doi [\mnras] {10.1093/mnrasl/sly093}, \href {https://ui.adsabs.harvard.edu/abs/2018MNRAS.479L..34M} {479, L34}

\bibitem[\protect\citeauthoryear{{Matthee} \& {Schaye}}{{Matthee} \& {Schaye}}{2019}]{matthee2019}
{Matthee} J.,  {Schaye} J.,  2019, \mn@doi [\mnras] {10.1093/mnras/stz030}, \href {https://ui.adsabs.harvard.edu/abs/2019MNRAS.484..915M} {484, 915}

\bibitem[\protect\citeauthoryear{{McDermid} et~al.,}{{McDermid} et~al.}{2015}]{mcdermid2015}
{McDermid} R.~M.,  et~al., 2015, \mn@doi [\mnras] {10.1093/mnras/stv105}, \href {https://ui.adsabs.harvard.edu/abs/2015MNRAS.448.3484M} {448, 3484}

\bibitem[\protect\citeauthoryear{{Moll{\'a}}, {V{\'\i}lchez}, {Gavil{\'a}n}  \& {D{\'\i}az}}{{Moll{\'a}} et~al.}{2006}]{molla2006}
{Moll{\'a}} M.,  {V{\'\i}lchez} J.~M.,  {Gavil{\'a}n} M.,   {D{\'\i}az} A.~I.,  2006, \mn@doi [\mnras] {10.1111/j.1365-2966.2006.10892.x}, \href {https://ui.adsabs.harvard.edu/abs/2006MNRAS.372.1069M} {372, 1069}

\bibitem[\protect\citeauthoryear{{Mu{\~n}oz} \& {Peeples}}{{Mu{\~n}oz} \& {Peeples}}{2015}]{munoz2015}
{Mu{\~n}oz} J.~A.,  {Peeples} M.~S.,  2015, \mn@doi [\mnras] {10.1093/mnras/stv048}, \href {https://ui.adsabs.harvard.edu/abs/2015MNRAS.448.1430M} {448, 1430}

\bibitem[\protect\citeauthoryear{{Naiman} et~al.,}{{Naiman} et~al.}{2018}]{tng2}
{Naiman} J.~P.,  et~al., 2018, \mn@doi [\mnras] {10.1093/mnras/sty618}, \href {https://ui.adsabs.harvard.edu/abs/2018MNRAS.477.1206N} {477, 1206}

\bibitem[\protect\citeauthoryear{{Nakajima}, {Ouchi}, {Isobe}, {Harikane}, {Zhang}, {Ono}, {Umeda}  \& {Oguri}}{{Nakajima} et~al.}{2023}]{nakajima2023}
{Nakajima} K.,  {Ouchi} M.,  {Isobe} Y.,  {Harikane} Y.,  {Zhang} Y.,  {Ono} Y.,  {Umeda} H.,   {Oguri} M.,  2023, \mn@doi [\apjs] {10.3847/1538-4365/acd556}, \href {https://ui.adsabs.harvard.edu/abs/2023ApJS..269...33N} {269, 33}

\bibitem[\protect\citeauthoryear{{Nelson} et~al.,}{{Nelson} et~al.}{2018}]{tng4}
{Nelson} D.,  et~al., 2018, \mn@doi [\mnras] {10.1093/mnras/stx3040}, \href {https://ui.adsabs.harvard.edu/abs/2018MNRAS.475..624N} {475, 624}

\bibitem[\protect\citeauthoryear{{Nelson} et~al.,}{{Nelson} et~al.}{2019}]{tng6}
{Nelson} D.,  et~al., 2019, \mn@doi [\mnras] {10.1093/mnras/stz2306}, \href {https://ui.adsabs.harvard.edu/abs/2019MNRAS.490.3234N} {490, 3234}

\bibitem[\protect\citeauthoryear{{Nersesian} et~al.,}{{Nersesian} et~al.}{2025}]{nersesian2025}
{Nersesian} A.,  et~al., 2025, \mn@doi [\aap] {10.1051/0004-6361/202452662}, \href {https://ui.adsabs.harvard.edu/abs/2025A&A...695A..86N} {695, A86}

\bibitem[\protect\citeauthoryear{{Nicholls}, {Sutherland}, {Dopita}, {Kewley}  \& {Groves}}{{Nicholls} et~al.}{2017}]{nicholls2017}
{Nicholls} D.~C.,  {Sutherland} R.~S.,  {Dopita} M.~A.,  {Kewley} L.~J.,   {Groves} B.~A.,  2017, \mn@doi [\mnras] {10.1093/mnras/stw3235}, \href {https://ui.adsabs.harvard.edu/abs/2017MNRAS.466.4403N} {466, 4403}

\bibitem[\protect\citeauthoryear{{O'Donnell}}{{O'Donnell}}{1994}]{odonnell1994}
{O'Donnell} J.~E.,  1994, \mn@doi [\apj] {10.1086/173713}, \href {https://ui.adsabs.harvard.edu/abs/1994ApJ...422..158O} {422, 158}

\bibitem[\protect\citeauthoryear{{Osterbrock} \& {Pogge}}{{Osterbrock} \& {Pogge}}{1985}]{osterbrock1985}
{Osterbrock} D.~E.,  {Pogge} R.~W.,  1985, \mn@doi [\apj] {10.1086/163513}, \href {https://ui.adsabs.harvard.edu/abs/1985ApJ...297..166O} {297, 166}

\bibitem[\protect\citeauthoryear{{Peeples} \& {Somerville}}{{Peeples} \& {Somerville}}{2013}]{peeples2013}
{Peeples} M.~S.,  {Somerville} R.~S.,  2013, \mn@doi [\mnras] {10.1093/mnras/sts158}, \href {https://ui.adsabs.harvard.edu/abs/2013MNRAS.428.1766P} {428, 1766}

\bibitem[\protect\citeauthoryear{{Peng}, {Maiolino}  \& {Cochrane}}{{Peng} et~al.}{2015}]{peng2015}
{Peng} Y.,  {Maiolino} R.,   {Cochrane} R.,  2015, \mn@doi [\nat] {10.1038/nature14439}, \href {https://ui.adsabs.harvard.edu/abs/2015Natur.521..192P} {521, 192}

\bibitem[\protect\citeauthoryear{{P{\'e}rez-Montero} et~al.,}{{P{\'e}rez-Montero} et~al.}{2013}]{pm2013}
{P{\'e}rez-Montero} E.,  et~al., 2013, \mn@doi [\aap] {10.1051/0004-6361/201220070}, \href {https://ui.adsabs.harvard.edu/abs/2013A&A...549A..25P} {549, A25}

\bibitem[\protect\citeauthoryear{{P{\'e}rez-Montero} et~al.,}{{P{\'e}rez-Montero} et~al.}{2016}]{pm2016}
{P{\'e}rez-Montero} E.,  et~al., 2016, \mn@doi [\aap] {10.1051/0004-6361/201628601}, \href {https://ui.adsabs.harvard.edu/abs/2016A\% 26A...595A..62P} {595, A62}

\bibitem[\protect\citeauthoryear{{Perez}, {Tissera}  \& {Blaizot}}{{Perez} et~al.}{2009}]{perez2009}
{Perez} J.,  {Tissera} P.,   {Blaizot} J.,  2009, \mn@doi [\mnras] {10.1111/j.1365-2966.2009.15033.x}, \href {https://ui.adsabs.harvard.edu/abs/2009MNRAS.397..748P} {397, 748}

\bibitem[\protect\citeauthoryear{{Pillepich} et~al.,}{{Pillepich} et~al.}{2018}]{tng1}
{Pillepich} A.,  et~al., 2018, \mn@doi [\mnras] {10.1093/mnras/stx3112}, \href {https://ui.adsabs.harvard.edu/abs/2018MNRAS.475..648P} {475, 648}

\bibitem[\protect\citeauthoryear{{Pillepich} et~al.,}{{Pillepich} et~al.}{2019}]{tng7}
{Pillepich} A.,  et~al., 2019, \mn@doi [\mnras] {10.1093/mnras/stz2338}, \href {https://ui.adsabs.harvard.edu/abs/2019MNRAS.490.3196P} {490, 3196}

\bibitem[\protect\citeauthoryear{{Pilyugin}, {V{\'\i}lchez}  \& {Thuan}}{{Pilyugin} et~al.}{2010}]{pilyugin2010}
{Pilyugin} L.~S.,  {V{\'\i}lchez} J.~M.,   {Thuan} T.~X.,  2010, \mn@doi [\apj] {10.1088/0004-637X/720/2/1738}, \href {https://ui.adsabs.harvard.edu/abs/2010ApJ...720.1738P} {720, 1738}

\bibitem[\protect\citeauthoryear{{Pistis} et~al.,}{{Pistis} et~al.}{2024}]{pistis2024}
{Pistis} F.,  et~al., 2024, \mn@doi [\aap] {10.1051/0004-6361/202346943}, \href {https://ui.adsabs.harvard.edu/abs/2024A&A...683A.203P} {683, A203}

\bibitem[\protect\citeauthoryear{{Salim}, {Lee}, {Ly}, {Brinchmann}, {Dav{\'e}}, {Dickinson}, {Salzer}  \& {Charlot}}{{Salim} et~al.}{2014}]{salim2014}
{Salim} S.,  {Lee} J.~C.,  {Ly} C.,  {Brinchmann} J.,  {Dav{\'e}} R.,  {Dickinson} M.,  {Salzer} J.~J.,   {Charlot} S.,  2014, \mn@doi [\apj] {10.1088/0004-637X/797/2/126}, \href {https://ui.adsabs.harvard.edu/abs/2014ApJ...797..126S} {797, 126}

\bibitem[\protect\citeauthoryear{{S{\'a}nchez Almeida} \& {Dalla Vecchia}}{{S{\'a}nchez Almeida} \& {Dalla Vecchia}}{2018}]{sm2018a}
{S{\'a}nchez Almeida} J.,  {Dalla Vecchia} C.,  2018, \mn@doi [\apj] {10.3847/1538-4357/aac086}, \href {https://ui.adsabs.harvard.edu/abs/2018ApJ...859..109S} {859, 109}

\bibitem[\protect\citeauthoryear{{S{\'a}nchez-Menguiano}, {S{\'a}nchez Almeida}, {Mu{\~n}oz-Tu{\~n}{\'o}n}  \& {S{\'a}nchez}}{{S{\'a}nchez-Menguiano} et~al.}{2020}]{sm2020}
{S{\'a}nchez-Menguiano} L.,  {S{\'a}nchez Almeida} J.,  {Mu{\~n}oz-Tu{\~n}{\'o}n} C.,   {S{\'a}nchez} S.~F.,  2020, \mn@doi [\apj] {10.3847/1538-4357/abba7c}, \href {https://ui.adsabs.harvard.edu/abs/2020ApJ...903...52S} {903, 52}

\bibitem[\protect\citeauthoryear{{S{\'a}nchez-Menguiano}, {S{\'a}nchez Almeida}, {S{\'a}nchez}  \& {Mu{\~n}oz-Tu{\~n}{\'o}n}}{{S{\'a}nchez-Menguiano} et~al.}{2024a}]{sm2024}
{S{\'a}nchez-Menguiano} L.,  {S{\'a}nchez Almeida} J.,  {S{\'a}nchez} S.~F.,   {Mu{\~n}oz-Tu{\~n}{\'o}n} C.,  2024a, \mn@doi [\aap] {10.1051/0004-6361/202346708}, \href {https://ui.adsabs.harvard.edu/abs/2024A&A...681A.121S} {681, A121}

\bibitem[\protect\citeauthoryear{{S{\'a}nchez-Menguiano}, {S{\'a}nchez}, {S{\'a}nchez Almeida}  \& {Mu{\~n}oz-Tu{\~n}{\'o}n}}{{S{\'a}nchez-Menguiano} et~al.}{2024b}]{sm2024a}
{S{\'a}nchez-Menguiano} L.,  {S{\'a}nchez} S.~F.,  {S{\'a}nchez Almeida} J.,   {Mu{\~n}oz-Tu{\~n}{\'o}n} C.,  2024b, \mn@doi [\aap] {10.1051/0004-6361/202348423}, \href {https://ui.adsabs.harvard.edu/abs/2024A&A...682L..11S} {682, L11}

\bibitem[\protect\citeauthoryear{{S{\'a}nchez} et~al.,}{{S{\'a}nchez} et~al.}{2013}]{sanchez2013}
{S{\'a}nchez} S.~F.,  et~al., 2013, \mn@doi [\aap] {10.1051/0004-6361/201220669}, \href {https://ui.adsabs.harvard.edu/abs/2013A&A...554A..58S} {554, A58}

\bibitem[\protect\citeauthoryear{{S{\'a}nchez} et~al.,}{{S{\'a}nchez} et~al.}{2016a}]{sanchez2016}
{S{\'a}nchez} S.~F.,  et~al., 2016a, \rmxaa, \href {https://ui.adsabs.harvard.edu/abs/2016RMxAA..52...21S} {52, 21}

\bibitem[\protect\citeauthoryear{{S{\'a}nchez} et~al.,}{{S{\'a}nchez} et~al.}{2016b}]{sanchez2016b}
{S{\'a}nchez} S.~F.,  et~al., 2016b, \rmxaa, \href {https://ui.adsabs.harvard.edu/abs/2016RMxAA..52..171S} {52, 171}

\bibitem[\protect\citeauthoryear{{S{\'a}nchez} et~al.,}{{S{\'a}nchez} et~al.}{2017}]{sanchez2017a}
{S{\'a}nchez} S.~F.,  et~al., 2017, \mn@doi [\mnras] {10.1093/mnras/stx808}, \href {https://ui.adsabs.harvard.edu/abs/2017MNRAS.469.2121S} {469, 2121}

\bibitem[\protect\citeauthoryear{{S{\'a}nchez} et~al.,}{{S{\'a}nchez} et~al.}{2019}]{sanchez2019}
{S{\'a}nchez} S.~F.,  et~al., 2019, \mn@doi [\mnras] {10.1093/mnras/sty2730}, \href {https://ui.adsabs.harvard.edu/abs/2019MNRAS.482.1557S} {482, 1557}

\bibitem[\protect\citeauthoryear{{S{\'a}nchez} et~al.,}{{S{\'a}nchez} et~al.}{2022}]{sanchez2022}
{S{\'a}nchez} S.~F.,  et~al., 2022, \mn@doi [\apjs] {10.3847/1538-4365/ac7b8f}, \href {https://ui.adsabs.harvard.edu/abs/2022ApJS..262...36S} {262, 36}

\bibitem[\protect\citeauthoryear{{Sanders} et~al.,}{{Sanders} et~al.}{2018}]{sanders2018}
{Sanders} R.~L.,  et~al., 2018, \mn@doi [\apj] {10.3847/1538-4357/aabcbd}, \href {https://ui.adsabs.harvard.edu/abs/2018ApJ...858...99S} {858, 99}

\bibitem[\protect\citeauthoryear{{Schaefer} et~al.,}{{Schaefer} et~al.}{2022}]{schaefer2022}
{Schaefer} A.~L.,  et~al., 2022, \mn@doi [\apj] {10.3847/1538-4357/ac651a}, \href {https://ui.adsabs.harvard.edu/abs/2022ApJ...930..160S} {930, 160}

\bibitem[\protect\citeauthoryear{{Schaye} et~al.,}{{Schaye} et~al.}{2015}]{schaye2015}
{Schaye} J.,  et~al., 2015, \mn@doi [\mnras] {10.1093/mnras/stu2058}, \href {https://ui.adsabs.harvard.edu/abs/2015MNRAS.446..521S} {446, 521}

\bibitem[\protect\citeauthoryear{{Schlegel}, {Finkbeiner}  \& {Davis}}{{Schlegel} et~al.}{1998}]{schlegel1998}
{Schlegel} D.~J.,  {Finkbeiner} D.~P.,   {Davis} M.,  1998, \mn@doi [\apj] {10.1086/305772}, \href {http://adsabs.harvard.edu/abs/1998ApJ...500..525S} {500, 525}

\bibitem[\protect\citeauthoryear{{Schmidt}}{{Schmidt}}{1963}]{schmidt1963}
{Schmidt} M.,  1963, \mn@doi [\apj] {10.1086/147553}, \href {https://ui.adsabs.harvard.edu/abs/1963ApJ...137..758S} {137, 758}

\bibitem[\protect\citeauthoryear{{Scholte} et~al.,}{{Scholte} et~al.}{2024}]{scholte2024}
{Scholte} D.,  et~al., 2024, \mn@doi [\mnras] {10.1093/mnras/stae2477}, \href {https://ui.adsabs.harvard.edu/abs/2024MNRAS.535.2341S} {535, 2341}

\bibitem[\protect\citeauthoryear{{Scott} et~al.,}{{Scott} et~al.}{2009}]{scott2009}
{Scott} N.,  et~al., 2009, \mn@doi [\mnras] {10.1111/j.1365-2966.2009.15275.x}, \href {https://ui.adsabs.harvard.edu/abs/2009MNRAS.398.1835S} {398, 1835}

\bibitem[\protect\citeauthoryear{{Scott} et~al.,}{{Scott} et~al.}{2017}]{scott2017}
{Scott} N.,  et~al., 2017, \mn@doi [\mnras] {10.1093/mnras/stx2166}, \href {https://ui.adsabs.harvard.edu/abs/2017MNRAS.472.2833S} {472, 2833}

\bibitem[\protect\citeauthoryear{{Setton} et~al.,}{{Setton} et~al.}{2022}]{setton2022}
{Setton} D.~J.,  et~al., 2022, \mn@doi [\apj] {10.3847/1538-4357/ac6096}, \href {https://ui.adsabs.harvard.edu/abs/2022ApJ...931...51S} {931, 51}

\bibitem[\protect\citeauthoryear{{Sharma}, {Masters}, {Stark}, {Garland}, {Drory}, {Fraser-McKelvie}  \& {Weijmans}}{{Sharma} et~al.}{2023}]{sharma2023}
{Sharma} A.,  {Masters} K.~L.,  {Stark} D.~V.,  {Garland} J.,  {Drory} N.,  {Fraser-McKelvie} A.,   {Weijmans} A.-M.,  2023, \mn@doi [\mnras] {10.1093/mnras/stad2695}, \href {https://ui.adsabs.harvard.edu/abs/2023MNRAS.526.1573S} {526, 1573}

\bibitem[\protect\citeauthoryear{{Simard}, {Mendel}, {Patton}, {Ellison}  \& {McConnachie}}{{Simard} et~al.}{2011}]{simard2011}
{Simard} L.,  {Mendel} J.~T.,  {Patton} D.~R.,  {Ellison} S.~L.,   {McConnachie} A.~W.,  2011, \mn@doi [\apjs] {10.1088/0067-0049/196/1/11}, \href {http://adsabs.harvard.edu/abs/2011ApJS..196...11S} {196, 11}

\bibitem[\protect\citeauthoryear{{Smee} et~al.,}{{Smee} et~al.}{2013}]{smee2013}
{Smee} S.~A.,  et~al., 2013, \mn@doi [\aj] {10.1088/0004-6256/146/2/32}, \href {http://adsabs.harvard.edu/abs/2013AJ....146...32S} {146, 32}

\bibitem[\protect\citeauthoryear{{Spitoni}, {Calura}, {Mignoli}, {Gilli}, {Silva Aguirre}  \& {Gallazzi}}{{Spitoni} et~al.}{2020}]{spitoni2020}
{Spitoni} E.,  {Calura} F.,  {Mignoli} M.,  {Gilli} R.,  {Silva Aguirre} V.,   {Gallazzi} A.,  2020, \mn@doi [\aap] {10.1051/0004-6361/202037879}, \href {https://ui.adsabs.harvard.edu/abs/2020A&A...642A.113S} {642, A113}

\bibitem[\protect\citeauthoryear{{Springel} et~al.,}{{Springel} et~al.}{2018}]{tng5}
{Springel} V.,  et~al., 2018, \mn@doi [\mnras] {10.1093/mnras/stx3304}, \href {https://ui.adsabs.harvard.edu/abs/2018MNRAS.475..676S} {475, 676}

\bibitem[\protect\citeauthoryear{{Stanton} et~al.,}{{Stanton} et~al.}{2024}]{stanton2024}
{Stanton} T.~M.,  et~al., 2024, \mn@doi [\mnras] {10.1093/mnras/stae1705}, \href {https://ui.adsabs.harvard.edu/abs/2024MNRAS.532.3102S} {532, 3102}

\bibitem[\protect\citeauthoryear{{Strom}, {Steidel}, {Rudie}, {Trainor}, {Pettini}  \& {Reddy}}{{Strom} et~al.}{2017}]{strom2017}
{Strom} A.~L.,  {Steidel} C.~C.,  {Rudie} G.~C.,  {Trainor} R.~F.,  {Pettini} M.,   {Reddy} N.~A.,  2017, \mn@doi [\apj] {10.3847/1538-4357/836/2/164}, \href {https://ui.adsabs.harvard.edu/abs/2017ApJ...836..164S} {836, 164}

\bibitem[\protect\citeauthoryear{{Telford}, {Dalcanton}, {Skillman}  \& {Conroy}}{{Telford} et~al.}{2016}]{telford2016}
{Telford} O.~G.,  {Dalcanton} J.~J.,  {Skillman} E.~D.,   {Conroy} C.,  2016, \mn@doi [\apj] {10.3847/0004-637X/827/1/35}, \href {https://ui.adsabs.harvard.edu/abs/2016ApJ...827...35T} {827, 35}

\bibitem[\protect\citeauthoryear{{Timmes}, {Woosley}  \& {Weaver}}{{Timmes} et~al.}{1995}]{timmes1995}
{Timmes} F.~X.,  {Woosley} S.~E.,   {Weaver} T.~A.,  1995, \mn@doi [\apjs] {10.1086/192172}, \href {https://ui.adsabs.harvard.edu/abs/1995ApJS...98..617T} {98, 617}

\bibitem[\protect\citeauthoryear{{Ting}, {Rix}, {Conroy}, {Ho}  \& {Lin}}{{Ting} et~al.}{2017}]{ting2017}
{Ting} Y.-S.,  {Rix} H.-W.,  {Conroy} C.,  {Ho} A. Y.~Q.,   {Lin} J.,  2017, \mn@doi [\apjl] {10.3847/2041-8213/aa921c}, \href {https://ui.adsabs.harvard.edu/abs/2017ApJ...849L...9T} {849, L9}

\bibitem[\protect\citeauthoryear{{Ting}, {Conroy}, {Rix}  \& {Asplund}}{{Ting} et~al.}{2018}]{ting2018}
{Ting} Y.-S.,  {Conroy} C.,  {Rix} H.-W.,   {Asplund} M.,  2018, \mn@doi [\apj] {10.3847/1538-4357/aac6c9}, \href {https://ui.adsabs.harvard.edu/abs/2018ApJ...860..159T} {860, 159}

\bibitem[\protect\citeauthoryear{{Torrey} et~al.,}{{Torrey} et~al.}{2018}]{torrey2018}
{Torrey} P.,  et~al., 2018, \mn@doi [\mnras] {10.1093/mnrasl/sly031}, \href {https://ui.adsabs.harvard.edu/abs/2018MNRAS.477L..16T} {477, L16}

\bibitem[\protect\citeauthoryear{{Torrey} et~al.,}{{Torrey} et~al.}{2019}]{torrey2019}
{Torrey} P.,  et~al., 2019, \mn@doi [\mnras] {10.1093/mnras/stz243}, \href {https://ui.adsabs.harvard.edu/abs/2019MNRAS.484.5587T} {484, 5587}

\bibitem[\protect\citeauthoryear{{Tortora}, {Hunt}  \& {Ginolfi}}{{Tortora} et~al.}{2022}]{tortora2022}
{Tortora} C.,  {Hunt} L.~K.,   {Ginolfi} M.,  2022, \mn@doi [\aap] {10.1051/0004-6361/202140414}, \href {https://ui.adsabs.harvard.edu/abs/2022A&A...657A..19T} {657, A19}

\bibitem[\protect\citeauthoryear{{Tremonti} et~al.,}{{Tremonti} et~al.}{2004}]{tremonti2004}
{Tremonti} C.~A.,  et~al., 2004, \mn@doi [\apj] {10.1086/423264}, \href {https://ui.adsabs.harvard.edu/abs/2004ApJ...613..898T} {613, 898}

\bibitem[\protect\citeauthoryear{{Trussler}, {Maiolino}, {Maraston}, {Peng}, {Thomas}, {Goddard}  \& {Lian}}{{Trussler} et~al.}{2020}]{trussler2020}
{Trussler} J.,  {Maiolino} R.,  {Maraston} C.,  {Peng} Y.,  {Thomas} D.,  {Goddard} D.,   {Lian} J.,  2020, \mn@doi [\mnras] {10.1093/mnras/stz3286}, \href {https://ui.adsabs.harvard.edu/abs/2020MNRAS.491.5406T} {491, 5406}

\bibitem[\protect\citeauthoryear{{Vale Asari}, {Couto}, {Cid Fernandes}, {Stasi{\'n}ska}, {de Amorim}, {Ruschel-Dutra}, {Werle}  \& {Florido}}{{Vale Asari} et~al.}{2019}]{valeasari2019}
{Vale Asari} N.,  {Couto} G.~S.,  {Cid Fernandes} R.,  {Stasi{\'n}ska} G.,  {de Amorim} A.~L.,  {Ruschel-Dutra} D.,  {Werle} A.,   {Florido} T.~Z.,  2019, \mn@doi [\mnras] {10.1093/mnras/stz2470}, \href {https://ui.adsabs.harvard.edu/abs/2019MNRAS.489.4721V} {489, 4721}

\bibitem[\protect\citeauthoryear{{Vaughan} et~al.,}{{Vaughan} et~al.}{2022}]{vaughan2022}
{Vaughan} S.~P.,  et~al., 2022, \mn@doi [\mnras] {10.1093/mnras/stac2304}, \href {https://ui.adsabs.harvard.edu/abs/2022MNRAS.516.2971V} {516, 2971}

\bibitem[\protect\citeauthoryear{{Veilleux} \& {Osterbrock}}{{Veilleux} \& {Osterbrock}}{1987}]{veilleux1987}
{Veilleux} S.,  {Osterbrock} D.~E.,  1987, \mn@doi [\apjs] {10.1086/191166}, \href {http://adsabs.harvard.edu/abs/1987ApJS...63..295V} {63, 295}

\bibitem[\protect\citeauthoryear{{Vera}, {Alonso}  \& {Coldwell}}{{Vera} et~al.}{2016}]{vera2016}
{Vera} M.,  {Alonso} S.,   {Coldwell} G.,  2016, \mn@doi [\aap] {10.1051/0004-6361/201628750}, \href {https://ui.adsabs.harvard.edu/abs/2016A&A...595A..63V} {595, A63}

\bibitem[\protect\citeauthoryear{{Vincenzo}, {Belfiore}, {Maiolino}, {Matteucci}  \& {Ventura}}{{Vincenzo} et~al.}{2016}]{vincenzo2016}
{Vincenzo} F.,  {Belfiore} F.,  {Maiolino} R.,  {Matteucci} F.,   {Ventura} P.,  2016, \mn@doi [\mnras] {10.1093/mnras/stw532}, \href {https://ui.adsabs.harvard.edu/abs/2016MNRAS.458.3466V} {458, 3466}

\bibitem[\protect\citeauthoryear{{Wake} et~al.,}{{Wake} et~al.}{2017}]{wake2017}
{Wake} D.~A.,  et~al., 2017, \mn@doi [\aj] {10.3847/1538-3881/aa7ecc}, \href {http://adsabs.harvard.edu/abs/2017AJ....154...86W} {154, 86}

\bibitem[\protect\citeauthoryear{{Walcher} et~al.,}{{Walcher} et~al.}{2019}]{walcher2019}
{Walcher} C.~J.,  et~al., 2019, \mn@doi [The Messenger] {10.18727/0722-6691/5118}, \href {https://ui.adsabs.harvard.edu/abs/2019Msngr.175...12W} {175, 12}

\bibitem[\protect\citeauthoryear{{Werle}, {Cid Fernandes}, {Vale Asari}, {Bruzual}, {Charlot}, {Gonzalez Delgado}  \& {Herpich}}{{Werle} et~al.}{2019}]{werle2019}
{Werle} A.,  {Cid Fernandes} R.,  {Vale Asari} N.,  {Bruzual} G.,  {Charlot} S.,  {Gonzalez Delgado} R.,   {Herpich} F.~R.,  2019, \mn@doi [\mnras] {10.1093/mnras/sty3264}, \href {https://ui.adsabs.harvard.edu/abs/2019MNRAS.483.2382W} {483, 2382}

\bibitem[\protect\citeauthoryear{{Westfall} et~al.,}{{Westfall} et~al.}{2019}]{westfall2019}
{Westfall} K.~B.,  et~al., 2019, \mn@doi [\aj] {10.3847/1538-3881/ab44a2}, \href {https://ui.adsabs.harvard.edu/abs/2019AJ....158..231W} {158, 231}

\bibitem[\protect\citeauthoryear{{Wild} et~al.,}{{Wild} et~al.}{2020}]{wild2020}
{Wild} V.,  et~al., 2020, \mn@doi [\mnras] {10.1093/mnras/staa674}, \href {https://ui.adsabs.harvard.edu/abs/2020MNRAS.494..529W} {494, 529}

\bibitem[\protect\citeauthoryear{{Wuyts} et~al.,}{{Wuyts} et~al.}{2011}]{wuyts2011}
{Wuyts} S.,  et~al., 2011, \mn@doi [\apj] {10.1088/0004-637X/742/2/96}, \href {https://ui.adsabs.harvard.edu/abs/2011ApJ...742...96W} {742, 96}

\bibitem[\protect\citeauthoryear{{Yan} et~al.,}{{Yan} et~al.}{2016a}]{yan2016a}
{Yan} R.,  et~al., 2016a, \mn@doi [\aj] {10.3847/0004-6256/151/1/8}, \href {http://adsabs.harvard.edu/abs/2016AJ....151....8Y} {151, 8}

\bibitem[\protect\citeauthoryear{{Yan} et~al.,}{{Yan} et~al.}{2016b}]{yan2016b}
{Yan} R.,  et~al., 2016b, \mn@doi [\aj] {10.3847/0004-6256/152/6/197}, \href {http://adsabs.harvard.edu/abs/2016AJ....152..197Y} {152, 197}

\bibitem[\protect\citeauthoryear{{Yan} et~al.,}{{Yan} et~al.}{2019}]{yan2019}
{Yan} R.,  et~al., 2019, \mn@doi [\apj] {10.3847/1538-4357/ab3ebc}, \href {https://ui.adsabs.harvard.edu/abs/2019ApJ...883..175Y} {883, 175}

\bibitem[\protect\citeauthoryear{{Yang}, {Scholte}  \& {Saintonge}}{{Yang} et~al.}{2024}]{yang2024}
{Yang} N.,  {Scholte} D.,   {Saintonge} A.,  2024, \mn@doi [\mnras] {10.1093/mnras/stad3917}, \href {https://ui.adsabs.harvard.edu/abs/2024MNRAS.52711043Y} {527, 11043}

\bibitem[\protect\citeauthoryear{{Zabludoff}, {Zaritsky}, {Lin}, {Tucker}, {Hashimoto}, {Shectman}, {Oemler}  \& {Kirshner}}{{Zabludoff} et~al.}{1996}]{zabludoff1996}
{Zabludoff} A.~I.,  {Zaritsky} D.,  {Lin} H.,  {Tucker} D.,  {Hashimoto} Y.,  {Shectman} S.~A.,  {Oemler} A.,   {Kirshner} R.~P.,  1996, \mn@doi [\apj] {10.1086/177495}, \href {https://ui.adsabs.harvard.edu/abs/1996ApJ...466..104Z} {466, 104}

\bibitem[\protect\citeauthoryear{{Zahid}, {Dima}, {Kudritzki}, {Kewley}, {Geller}, {Hwang}, {Silverman}  \& {Kashino}}{{Zahid} et~al.}{2014}]{zahid2014}
{Zahid} H.~J.,  {Dima} G.~I.,  {Kudritzki} R.-P.,  {Kewley} L.~J.,  {Geller} M.~J.,  {Hwang} H.~S.,  {Silverman} J.~D.,   {Kashino} D.,  2014, \mn@doi [\apj] {10.1088/0004-637X/791/2/130}, \href {https://ui.adsabs.harvard.edu/abs/2014ApJ...791..130Z} {791, 130}

\bibitem[\protect\citeauthoryear{{Zenocratti}, {De Rossi}, {Theuns}  \& {Lara-L{\'o}pez}}{{Zenocratti} et~al.}{2022}]{zenocratti2022}
{Zenocratti} L.~J.,  {De Rossi} M.~E.,  {Theuns} T.,   {Lara-L{\'o}pez} M.~A.,  2022, \mn@doi [\mnras] {10.1093/mnras/stac906}, \href {https://ui.adsabs.harvard.edu/abs/2022MNRAS.512.6164Z} {512, 6164}

\bibitem[\protect\citeauthoryear{{Zhou} et~al.,}{{Zhou} et~al.}{2022}]{zhou2022}
{Zhou} Y.,  et~al., 2022, \mn@doi [\mnras] {10.1093/mnras/stac2016}, \href {https://ui.adsabs.harvard.edu/abs/2022MNRAS.515.5081Z} {515, 5081}

\bibitem[\protect\citeauthoryear{{Zhuang} et~al.,}{{Zhuang} et~al.}{2024}]{zhuang2024}
{Zhuang} Z.,  et~al., 2024, \mn@doi [\apj] {10.3847/1538-4357/ad5ff8}, \href {https://ui.adsabs.harvard.edu/abs/2024ApJ...972..182Z} {972, 182}

\bibitem[\protect\citeauthoryear{{Zinchenko} \& {V{\'\i}lchez}}{{Zinchenko} \& {V{\'\i}lchez}}{2024}]{zinchenko2024}
{Zinchenko} I.~A.,  {V{\'\i}lchez} J.~M.,  2024, \mn@doi [\aap] {10.1051/0004-6361/202450273}, \href {https://ui.adsabs.harvard.edu/abs/2024A&A...690A..13Z} {690, A13}

\bibitem[\protect\citeauthoryear{{de Jong} et~al.,}{{de Jong} et~al.}{2019}]{dejong2019}
{de Jong} R.~S.,  et~al., 2019, \mn@doi [The Messenger] {10.18727/0722-6691/5117}, \href {https://ui.adsabs.harvard.edu/abs/2019Msngr.175....3D} {175, 3}

\bibitem[\protect\citeauthoryear{{van der Wel} et~al.,}{{van der Wel} et~al.}{2014}]{vdw2014}
{van der Wel} A.,  et~al., 2014, \mn@doi [\apj] {10.1088/0004-637X/788/1/28}, \href {https://ui.adsabs.harvard.edu/abs/2014ApJ...788...28V} {788, 28}

\makeatother
\end{thebibliography}

\label{lastpage}
\end{document}